\documentclass[aps,pre,superscriptaddress,showpacs,showkeys,reprint]{revtex4-1}
\usepackage{amsmath}
\usepackage{amsfonts}
\usepackage{mathrsfs}
\usepackage{amsthm}
\usepackage{graphicx}
\usepackage{subfigure}
\usepackage{amssymb}
\usepackage{hyperref}
\usepackage{color}
\usepackage{url}
\usepackage[abs]{overpic}
\usepackage{epstopdf}
\usepackage[usenames,dvipsnames]{xcolor}
\usepackage{soul}
\usepackage{comment}
\usepackage{grfext}
\usepackage{grffile}

\PrependGraphicsExtensions*{.png,.PNG}

\def\kmax{k_{\rm max}}
\def\bk{{\bf k}}
\def\bu{{\bf u}}

\def\mO{{\mathcal O}}

\def \lf{{\ell_{\rm f}}}
\def \kf{{k_{\rm f}}}

\def\inj{{\rm inj}}
\def\eq{{\rm eq}}
\def\th{{\rm th}}

\newcommand{\Aone}{$\tt{B_{128}}$}
\newcommand{\Atwo}{$\tt{A_{128}}$}
\newcommand{\B}{$\tt{ A_{64}}$}

\newcommand{\Rr}{\mathcal{R}_{\rm r}}
\newcommand{\RL}{\mathcal{R}_{\rm L}}
\newcommand{\nur}{\nu_{\rm r}}

\newcommand{\tOm}{\tilde \Omega}

\def\av#1{{\left\langle #1\right\rangle}}  
\def\figref#1{Fig.~{\ref{#1}}}

\def\old{\color{black}}
\newcommand{\inphyniaddress}{Universit\'e C\^ote d'Azur, Institut de Physique de Nice (INPHYNI),
CNRS UMR 7010, Parc Valrose, 06108 Nice Cedex 2, France}
\newcommand{\impaaddress}{Instituto Nacional de Matem\'atica Pura e Aplicada, IMPA, 22460-320 Rio de Janeiro, Brazil}
\newcommand{\ocaaddress}{Universit\'e C\^ote d'Azur, CNRS, OCA, Laboratoire Lagrange, Bd. de l'Observatoire, Nice, France}
\newcommand{\ceaaddress}{DSM/IRAMIS/SPEC, CNRS UMR 3680, CEA, Universit\'e Paris-Saclay, 91190 Gif sur Yvette, France}

\begin{document}
\title{Phase transition in  time-reversible Navier-Stokes equations}

\author{Vishwanath Shukla}
\email{research.vishwanath@gmail.com}
\affiliation{Department of Physics, Indian Institute of Technology Kharagpur, Kharagpur 721302, India.}
\affiliation{\inphyniaddress}
\author{B\'ereng\`ere Dubrulle}
\affiliation{\ceaaddress}
\author{Sergey Nazarenko}
\affiliation{\inphyniaddress}
\author{Giorgio Krstulovic}
\affiliation{\ocaaddress}
\author{Simon Thalabard}
\email{simon.thalabard@impa.br}
\affiliation{\impaaddress}
\date{\today}

\begin{abstract}
We present a comprehensive study of the statistical features of a three-dimensional 
time-reversible  Navier-Stokes (RNS) system, wherein the standard viscosity $\nu$ is replaced 
by a fluctuating thermostat that dynamically compensates for fluctuations in the 
total energy. %
We analyze the statistical features of the RNS steady states in terms of a non-negative dimensionless control parameter 
$\Rr$, which quantifies the balance between the fluctuations of 
kinetic energy at the forcing length scale $\ell_{\rm f}$ and the total energy $E_0$.
For small $ \Rr$, the RNS equations are found to produce ``warm'' stationary statistics, \emph{e.g.} characterized by the  partial thermalization of the small length-scales. 
For large $\Rr$, the stationary solutions have features akin to standard hydrodynamic ones: They have  compact energy support in $k$-space and are essentially insensitive to  the truncation scale $\kmax$.  
The transition between the two statistical regimes is observed to be smooth but rather sharp.
Using insights from a diffusion model of turbulence (Leith model), we  argue that the transition is in fact akin to a \emph{continuous
phase transition}, where  $\Rr$ indeed behaves as a thermodynamic control parameter, \emph{e.g.} a temperature.  A relevant order-parameter can be suitably defined in terms of a (normalized)   enstrophy, while the symmetry breaking parameter $h$ is identified as (one over)  the truncation scale $\kmax$.
We find that the signatures of the phase transition close to the critical point $\Rr^\star$ can essentially be deduced from  a heuristic mean-field Landau free energy.
This point of view allows us to reinterpret the relevant asymptotics in which the  \textit{dynamical ensemble equivalence} conjectured by Gallavotti, \emph{Phys.Lett.A, 223, 1996} could hold true.  We argue that Gallavotti's limit is precisely the joint limit $\Rr \overset{>}{\to} \Rr^\star$ and $h\overset{>}{\to}  0$, with the overset symbol ``$>$'' indicating that these limits are approached from above. The limit therefore relates to the statistical features at the critical point.
In this regime, our numerics indicate that the low-order statistics of the 3D RNS  are indeed qualitatively similar to those observed in direct numerical simulations of the standard Navier-Stokes (NS) equations  with viscosity chosen so as to match the average value of the reversible viscosity.
This result suggests that Gallavotti's \textit{equivalence conjecture}  could indeed be of relevance to model 3D turbulent statistics, and provides a clear guideline for further numerical investigations at higher resolutions.
\old
\end{abstract}


\maketitle

\section{Introduction}
Describing the irreversible behaviors of macroscopic observables arising  from  time-reversible 
microscopic dynamics  is the central long-standing theme of non-equilibrium statistical
mechanics \cite{cercignani1988boltzmann,cercignani2013mathematical,chibbaro2014reductionism}.
When there exists a wide scale separation between the microscopic and the macroscopic scales, the emergence of irreversibility 
can in general be formalized using a variety of reduction techniques 
including but not limited to stochastic equations, diffusion   or projection operator formalisms that  model the collective evolution of the fast variables \cite{zwanzig1973nonlinear,
zwanzig2001nonequilibrium,turkington2013optimization,gardiner2009stochastic}.
The scope of many other promising strategies is still an active area of research~\cite{onsager1953fluctuations,bertini2002macroscopic,ottinger2005beyond,
kurchan2007non,kraaij2019fluctuation}; therefore, a systematic framework is lacking that allows to derive,
from first principles, a non-equilibrium thermodynamic formalism to account for the macroscopic irreversibility.

In the context of  three-dimensional (3D) stationary homogeneous isotropic turbulence, 
a hallmark of irreversibility is the phenomenon of anomalous dissipation, 
namely the fact that the rate of energy dissipation $\epsilon$ becomes 
finite as  the separation between the injection and the dissipative 
viscous scales become infinite. The breaking of detailed balance
is then made apparent through the celebrated four-fifth law (see, 
e.g. Ref.~\cite{frisch1995turbulence}), which ties $\epsilon$ to the average of the 
cube of the longitudinal velocity increments. 
This is an anomalous feature, as in the limit of vanishing viscosity (infinite 
Reynolds number) the flow could in principle formally be described by the time-symmetric
Euler equations.\\

A thorough description of irreversibility in turbulence requires to underpin its 
precise features and in recent years this problem has witnessed a renewed interest.
In particular, non-trivial signatures of irreversibility have been identified 
on the Lagrangian statistics: Both experiments and large numerical simulations 
have demonstrated that these depend on the forward-in-time or backward-in-time
conditioning~\cite{berg2006PRE,buaria2015PoF,buaria2016lagrangian,ray2018non}.
For instance, both fluid and heavy particles tend to gain kinetic energy slowly but  loose it
rapidly  along their Lagrangian trajectories ~\cite{XuPNAS2014,xu2016lagirrever,Bhatnagar18irrv}: This is a clear example of an irreversible behavior, whose origin is related
to the vortex stretching and generation of small length scales~\cite{PumirPRL2016}; 
this persists even in the limit of vanishing viscosity.\\

One important difficulty in studying turbulent irreversibility comes from its 
asymptotic nature. Even massive computational effort in numerically integrating 
the NS equations may fail in clearly disentangling the 
finite-Reynolds-number effects from its truly asymptotic features~\cite{Jucha2014PRL,iyer2018steep}.
An alternative approach is to modify the governing equations to make them time-reversible, and then study whether the irreversible signatures of turbulence are still present under suitably defined limits. 
An early example of such an approach is that of the ``constrained Euler system'' considered in 
Ref.~\cite{ShePRL1993}, wherein the energies contained within narrow wave number 
shells are held constant in time. The resulting system was shown to reproduce many
of the standard statistical features of  isotropic Navier-Stokes (NS) turbulence, including intermittency.

In Ref.~\cite{gallv96PLAequivconj}, another time-reversible governing equation was proposed, 
based on the assumption that the fluid is not subjected to the usual viscous dissipation,
but rather to a modified dissipation mechanism, obtained by imposing a global constraint on the system.
This results in a time-reversal invariant dissipative term characterized by the appearance of a 
``reversible viscosity'' that balances the energy injection by behaving like a ``thermostatting term'', while a prescribed macroscopic observable such as the total energy or the total enstrophy remains constant in time.
An equivalence between these time-reversible formulations and the standard
NS dynamics was postulated to hold true in the limit of high Reynolds number \cite{gallv96PLAequivconj}, as
a consequence of a more general \textit{equivalence of dynamical ensembles}
for non-equilibrium systems \cite{GallaCohenPRL95}.
If this ``equivalence conjecture'' is true, at least for suitable choices of thermostat, then the statistical 
features of turbulent flows in the inertial range can be obtained by adopting two distinct approaches, which model
microscopic dissipation differently, but yield an equivalent macroscopic behavior.

The use of the reversible formulation opens up the possibility to explore the 
implications of the chaotic hypothesis~\cite{gallv95chaotichypo} for the fluctuations of the
local observables and the Lyapunov spectrum. This perspective has motivated many investigations, 
including numerical~\cite{biferaleTimerev98,rondoniRevdisp2Dturb99,aumaitre2001power,
gallv2004lyapunov}  and experimental ones~\cite{ciliberto1998experimental}.

Numerical tests probing the equivalence of dynamical ensembles have been performed in various settings,
but so far only for simple models rather than the full 3D NS equations.
For instance, the time-reversible version of the shell model of turbulence obtained by imposing
a global constraint of energy conservation was investigated in~\cite{biferaleTimerev98}.
It was found that as the amplitude of the external force is varied,
from zero to high values, the system exhibits a smooth transition from an 
equilibrium state to a non-equilibrium stationary state with an energy cascade
from large to small scales. 

Such models have also been studied in combination with various kinds of thermostats. 
Recent results suggest that the relevance of the equivalence conjecture might crucially 
depend on which  macroscopic observable is chosen to be held constant~\cite{biferale2018equivconj,gallv2019private}.
Insights on how macroscopic irreversibility is linked to the non-equilibrium energy cascade process rather than to
the explicit breaking of the time-reversal invariance due to viscous  dissipation were also reported in Ref.~\cite{depietro2018shell}.


The validity of the equivalence conjecture along with various consequences of  the chaotic
hypothesis were tested for  incompressible two-dimensional (2D)
flows~\cite{rondoniRevdisp2Dturb99,gallv2004lyapunov}.
Direct numerical simulations (DNS) of the incompressible 2D NS
equations  were compared to their reversible counterpart, in order  to examine the fluctuations 
of  global quadratic quantities in statistically stationary states.
A comparative study of the Lyapunov spectra showed that they overlap~\cite{gallv2004lyapunov}. 
These studies naturally went beyond the reduced models of turbulence and dealt with
the full governing equations, though with small number
of Fourier modes, and provided an additional support in
favor of the conjecture.

The above discussion suggests that the (time) Reversible Navier-Stokes (RNS) systems,
as prescribed by the equivalence conjecture, can perhaps provide a generalized framework which is 
capable of producing genuine turbulent statistics arsing from a time-reversible dynamics.
This is especially useful for understanding the anomalous turbulent signatures.
Therefore, the recent works based on the shell models of turbulence~\cite{biferale2018equivconj,depietro2018shell}
are a step forward in understanding the Gallavotti's conjecture.

To the best our knowledge, no systematic attempt  has been made so far in order to clearly achieve 
the limit in which the equivalence conjecture could supposedly hold true,  \emph{e.g.} the 
limit $\nu \to 0$ for the \emph{full}  3D NS equations. 
The obvious reason for this, is the fact that this  question is both subtle and \emph{a priori} 
difficult to tackle from a numerical perspective. Any numerical scheme involves a cutoff 
scale $\kmax$, and the desired asymptotics is then necessarily   a joint limit 
$\kmax\to \infty,\, \nu \to 0$.  In principle, these two limits do not commute. 
In the context of Gallavotti's original equivalence conjecture, one should clearly 
let $\kmax \to \infty$ before letting $\nu \to 0$, and in our view even a 
phenomenological hint as to whether the equivalence conjecture should reasonably  
hold in this limit is perhaps currently lacking. 
To gain such an intuition,  one should probably first understand the nature of statistical 
regimes that the  RNS  dynamics is likely to generate. Yet, systematic overviews, to this day, 
at best are either essentially qualitative or simply absent, especially for the case of 3D RNS. 
The present paper intends to fill this gap.\\

Our work offers a comprehensive study 
of the statistical features of a 3D time-reversible NS
system, in which the standard viscosity is replaced by a fluctuating thermostat 
that dynamically compensates for fluctuations in the total energy.
To identify different statistical regimes of this system,
we introduce a non-negative dimensionless control parameter 
$\Rr=f_0 \ell_{\rm f}/E_0$, which quantifies the balance between the injection of 
kinetic energy at the forcing length-scale $\ell_{\rm f}$ and the total energy $E_0$.
We find that the system exhibits a smooth transition from a high-enstrophy,
truncation effects dominated phase at small $\Rr$ to low-enstrophy, 
hydrodynamical states at large $\Rr$ . 
This transition has features akin to a 
continuous phase transition, with average enstrophy as an order parameter.

For small values of $\Rr$, the RNS equations produce steady states that exhibit 
close-to-equilibrium 
Gibbs-type statistics at small length-scales. Following Ref.~\cite{connaughton2004warm}, 
we refer to such states as \emph{warm} states (solutions). The terminology is  simply meant 
to convey the idea
that the spectra being partially thermalized at the ultra-violet end, should behave akin to a heat-bath, 
a feature previously observed in truncated fluid models~\cite{cichowlas2005effective,krstulovic2008two,thalabard2016optimal}. 
For large $\Rr$, the stationary solutions have compact energy support in $k$-space and are found 
to be essentially insensitive to the cutoff scale $\kmax$ (later precisely defined) and we refer 
to these kind of states as being of \emph{hydrodynamic} type. 
Furthermore, using insights from a reversible non-linear diffusion model of turbulence (Leith model), 
we  argue that the transition is in fact akin to a \emph{continuous phase transition}, 
and that  $\Rr$ indeed behaves as a thermodynamic control parameter, \emph{e.g.} a temperature.  
Also, as mentioned above, a relevant order-parameter can be suitably defined in terms of a (normalized)   
enstrophy, while the symmetry breaking parameter $h$ is identified as (one over)  the truncation 
scale $\kmax$.
We find that the signatures of the phase transition close to the critical point $\Rr^\star$ can essentially be deduced from  a heuristic mean-field Landau free energy.
This point of view allows us to reinterpret the relevant asymptotics in which Gallavotti's  conjecture could hold true.  Gallavotti's limit precisely corresponds to the joint limit  $\Rr \overset{>}{\to} \Rr^\star$ and $h\overset{>}{\to}  0$, with overset ``$>$'' meaning that the critical point is approached from above. It  therefore relates to the statistics in the neighborhood of  the critical point.
In this regime, our numerics indicate that the 3D RNS  steady statistics mimic  their standard NS counterpart, with viscosity matching the average value of the reversible viscosity. This result  hints towards the validity of the equivalence conjecture.\\

The remainder of this paper is organized as follows. 
\S~\ref{sec:detailrevNS} introduces  the RNS equations and the control parameter $\Rr$.
We  schematically discuss the expected statistical features of the RNS states in the 
two opposite asymptotic limits: $\Rr \to 0$ and $\Rr \to \infty$.
\S~\ref{sec:methods}  describes  the outcomes of our RNS numerics, and presents a detailed overview of 
the different statistical regimes  which we observe. We identify  a small crossover range of $\Rr$,  
wherein the RNS states  continuously transits from being  ``warm'' to ``hydrodynamic''. 
\S~\ref{sec:Leith}   discusses insights obtained from the  analysis of a suitably defined ``reversible Leith model'', the statistical regimes of which are interpreted within the framework of a mean-field second-order Landau theory. 
\S~\ref{sec:RLtoRNS} extends the discussion to the RNS system, and reformulates   
the equivalence in a thermodynamic framework. We  compare low-order RNS and NS statistics at values of $\Rr$
that are slightly above the critical point, and argue that this is indeed the relevant regime to consider. 
\S~\ref{sec:conclusions} summarizes our findings and presents some perspectives.
\old

\section{The (time) Reversible Navier-Stokes equations}
\label{sec:detailrevNS}
\subsection{Formal definitions }
The spatio-temporal evolution of the velocity field $\mathbf{u}(\mathbf{x},t)$
describing an incompressible fluid flow within a spatial domain ${\mathcal D}$ is governed by the Navier-Stokes
(NS) equations
\begin{equation} \label{eq:NSeq}
\frac{\partial \mathbf{u}}{\partial t} + \left(\mathbf{u}\cdot\nabla\right)\mathbf{u}
= -\nabla p + \nu\nabla^2\mathbf{u} + \mathbf{f},
\end{equation}
where  $\nu$ is the kinematic viscosity, 
$p$ is the pressure field and $\mathbf{f}$ is the forcing term, acting at large length-scales,
to sustain a statistically steady state. The incompressibility is ensured by
requiring $\nabla\cdot\mathbf{u}=0$ and the fluid density is set to $1$.

In presence of the viscous dissipation term $\nu\nabla^2\mathbf{u}$,   
the resulting macroscopic dynamics is clearly irreversible, as the NS equations~\eqref{eq:NSeq}  
are not invariant under the transformation 
\begin{equation}
\mathcal{T}: \, t \to -t;\, \mathbf{u} \to -\mathbf{u}.
\end{equation}
We now follow Ref.~\cite{gallv96PLAequivconj}, and alter the dissipation operator
term to make it invariant under the
transformation $\mathcal{T}$. The modification consists  in transforming the 
dissipation operator into a  thermostat,  
so that a certain macroscopic quantity, such as  the total  energy or the total 
enstrophy, becomes a conserved quantity.
While Ref.~\cite{gallaPhysD97} discusses several implementations of this idea, 
we here choose to follow Ref.~\cite{rondoniRevdisp2Dturb99,biferaleTimerev98}, and 
impose a constraint on the total kinetic energy. 
An elementary calculation shows that in order for the energy to be held constant,  the 
viscosity must fluctuate as 
\begin{equation} \label{eq:revviscosity}
\nu_{\rm r}[\mathbf{u}] = \frac{\int_{\mathcal{D}}\,\mathbf{f}\cdot
\mathbf{u}\,d\mathbf{x}}
{\int_{\mathcal{D}}\,(\nabla\times\mathbf{u})^2\,d\mathbf{x}}.
\end{equation}

The ``reversible viscosity'' is a functional of $\mathbf{u}$ and depends on the state of the system.
We refer to the equations obtained by replacing the constant in time viscosity $\nu$ in the
NS equations~\eqref{eq:NSeq} with the state dependent $\nu_{\rm r}$ as the ``Reversible Navier-Stokes'' 
(RNS) equations:
\begin{equation} \label{eq:RNSeq}
\frac{\partial \mathbf{u}}{\partial t} + \left(\mathbf{u}\cdot\nabla\right)\mathbf{u}
= -\nabla p + \nu_{\rm r}\nabla^2\mathbf{u} + \mathbf{f},
\end{equation}
where we still enforce  incompressibility as $\nabla\cdot\mathbf{u}=0$.

\subsection{Control parameter $\Rr$}
\label{ssec:Rr}

To characterize the statistical steady states of the RNS system,
we use the dimensionless control parameter 
\begin{equation}\label{eq:dimlessnoRNS}
\mathcal{R}_{\rm r} = \frac{f_0 \ell_{\rm f}}{E_0},
\end{equation}
where $E_0$ is the total (conserved) energy fixed by the initial state, $f_0$ is the forcing amplitude 
and $\ell_{\rm f}$ is the energy injection length scale. 

Despite its suggestive name, the control parameter  $\Rr$ should not  
be interpreted as either a ``Reversible Reynolds number'' 
or an inverse thereof~: None of the two asymptotic regimes $\Rr \to 0 $  and $\Rr \to \infty$ 
describe a fully-developed turbulent state.
This is perhaps slightly counter-intuitive as when $\Rr \to 0$ the RNS dynamics formally reduces to the freely evolving Euler equations. 
There is yet no reason to expect this limit to produce a ``fully developed turbulent'' steady state, as it corresponds to a very specific \emph{joint limit}, where  both the viscous and the forcing  term simultaneously vanish.  Fully developed turbulence is in principle generated from the  NS equations in a different manner,  that is by letting the standard viscosity $\nu \to 0$ at fixed value of the forcing $f_0$ \cite{frisch1995turbulence}. There is therefore no reason that both limits coincide.

The present work relies on numerical integration of the RNS equations. 
The fact that the limit $\Rr \to 0$ is not linked with the fully-developed turbulence becomes clear 
from the following arguments. 
Indeed, any numerical calculation  involves a finite resolution, or equivalently a finite number of degrees of
freedom.  At fixed resolution,  the limit $\Rr \to 0$ does not yield the Euler equations but rather their
\emph{truncated} counterpart. 
Hence, the  numerical integration of the truncated RNS equations, in the limit $\Rr\to0$, will converge towards an \emph{absolute equilibrium equipartition state} that corresponds to the equipartition of energy among the modes;
this state has Gaussian statistics quite unlike a fully developed turbulent state.
We  refer the reader to Appendix~\ref{sec:abseq} for further details on absolute equilibria and 
truncated Euler flows.

The limit $\Rr \to \infty$, on the contrary, resembles an over-damped dynamics: 
In this limit, the forcing is infinitely large compared to the energy retained in the system. 
Therefore, any energy injected at length-scale $\lf$ should in principle be immediately removed by
the reversible viscosity, thereby suppressing the  nonlinear transfer of energy. 
This asymptotic steady state is mostly insensitive to the  number of modes used in 
the numerical simulations.

\begin{figure}
\includegraphics[width=\columnwidth,trim=0cm 1cm 0cm 1cm, clip]{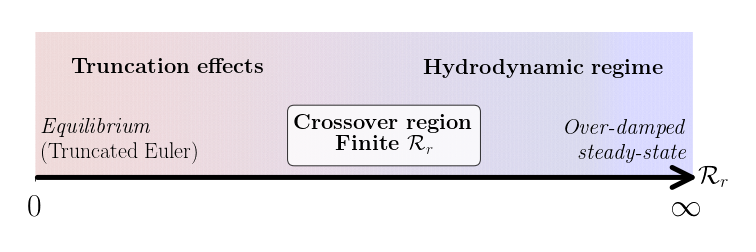}
\caption{A schematic illustration of the phase diagram of the (truncated) RNS system.
	The small $\Rr$ truncation dominated regime and the large $\Rr$ hydrodynamic regimes are separated 
	by a crossover region, in accordance with the heuristic description of \S~\ref{ssec:Rr} based on the two
	asymptotic 	limits $\Rr \to 0 $ and $\Rr \to \infty$.}
\label{fig:SketchI}
\end{figure}

The two asymptotic phases should obviously cross over  at intermediate values of $\Rr$, and this is very
schematically summarized by the  diagram sketched in 
\figref{fig:SketchI}. Our numerical simulations intend to substantiate this crude phenomenological 
overview, and in particular  provide a detailed characterization of the RNS statistical regimes when $\Rr$ takes a finite value.

\section{Numerical experiments}

\label{sec:methods}

We begin this section with a brief overview of our numerical methods that we use to study the
RNS system, followed by a comprehensive description of the results obtained from the numerical simulations.
We show that the phase diagram depicted in \figref{fig:SketchI} is correct. 
The RNS system indeed has two distinct statistical regimes separated by a crossover region; the transition 
between these two regimes has the character of a continuous-phase transition.


\subsection{Details of the simulations}
\subsubsection{Numerical schemes}
\label{sec:implementations}

We perform numerical simulations of the 3D NS Eqs.~\eqref{eq:NSeq} and 
the 3D RNS Eqs.~\eqref{eq:RNSeq} by using the Fourier pseudo-spectral method, implemented
in an efficient, parallel numerical code VIKSHOBHA~\cite{DebuePRE18}.

The velocity field $\mathbf{u}$ is solved inside a cubic domain $\mathcal D$ of side $2\pi$, and 
is prescribed to be triply-periodic. Therefore, it can be represented by the Fourier series
\begin{equation*}
\mathbf{u}(\mathbf{x},t) = \sum_{\bk} \hat{\bu} (\bk,t) 
\exp{(i\bk \cdot\mathbf{x})},
\end{equation*}
where $\mathbf{k}=(k_1,k_2,k_3)$, $k_i\in [-N_c/2,N_c/2-1]$ represent the 3D wave vectors and
$N_c$ is the number of collocation points.
The incompressibility condition is used to eliminate the pressure term by introducing 
a transverse projection operator $\mathbb{P}_{\rm i,j}(\mathbf{k})=\delta_{\rm i,j}-k_{\rm i}k_{\rm j}/k^2$ 
that projects the nonlinear term on a plane perpendicular to $\mathbf{k}$.
The Fourier pseudo-spectral method relies on the computation of the linear terms in Fourier space 
and the nonlinear terms in real space, before transforming them back to Fourier space.
Aliasing errors are removed using the standard  
$2/3$-dealiasing rule, so that the maximum wave number in our simulations is $\kmax =N_c/3$. 
Both the NS and the RNS  dynamics are evolved in time using a second-order Runge-Kutta scheme. 
In our RNS numerics,  the time-step was kept very small $dt=7.5 \times 10^{-4}$, and this allowed  us to have a 
very accurate conservation of the energy; errors are below 0.03 percent in our runs.
\old
\subsubsection{Initial data and forcing}
Both the RNS and NS runs  are initiated from the following  Taylor-Green velocity field:
\begin{equation*}
\begin{split}
	u_{x} &= u_0\,\sin(x)\cos(y)\cos(z),\\
	u_{y} &=-u_0\,\cos(x)\sin(y)\cos(z),\\
	u_{z} &= 0,
\end{split}
\end{equation*}
where the   coefficient $u_0$ sets the value of the initial energy.

In order to obtain statistically steady states, we inject energy in the system by using the Taylor-Green forcing:
\begin{equation*}
\begin{split}
	f_{x} &= f_0\sin(\tilde{k}_{\rm f} x)\cos(\tilde{k}_{\rm f} y)\cos(\tilde{k}_{\rm f} z),\\
	f_{y} &=-f_0\cos(\tilde{k}_{\rm f} x)\sin(\tilde{k}_{\rm f} y)\cos(\tilde{k}_{\rm f} z),\\
	f_{z} &= 0,
\end{split}
\end{equation*}
where $f_0$ and $\tilde{k}_{\rm f}$  are respectively the forcing amplitude and wave number.
We write $k_{\rm f}:=\sqrt{3}\tilde{k}_{\rm f}$ as the norm of the forcing wave vector 
$\mathbf{k}_{\rm f}=\left(\tilde{k}_{\rm f},\tilde{k}_{\rm f},\tilde{k}_{\rm f}\right)$.

As an aside, let us recall that the   Taylor-Green flow has a vanishing total helicity, \emph{e.g.} 
$\int_{\mathcal D} \mathbf{u}.(\nabla \times \mathbf{u}) =0$.

\subsubsection{Conventional definitions}
\label{sec:conventions}
We compute the isotropic energy spectrum as
\begin{equation*}
E(k,t) :=\frac{1}{2}\hspace{-8pt}\sum_{\substack{\bk:\\\rm k-\frac{1}{2}<|\bk|\leq k+\frac{1}{2}}}
\hspace{-12pt} |\hat{\mathbf{u}}(\bk,t)|^2,
\end{equation*}
from which both the (total) energy  $E := \sum_{k=1}^{\kmax}E(k,t)$  and the
enstrophy $\Omega := \sum_{k=1}^{\kmax}k^2E(k,t)$  are estimated.\\

The non-linear energy fluxes are defined through 
\begin{equation*}
	\begin{split}
		 &\Pi(k,t) = \sum_{|{\bf k}| \ge k} T({\bf k},t), \text{\;\;where}\\
	T(\bk,t ) & := \Re\left\{\sum_{\substack{ i,j=1,2,3}} \hspace{-6pt}\hat{u}^{\star}_{i}[\bk,t]\,\,%
	\mathbb{P}_{i,j} (\bk)%
	\left[\widehat{\bu \times (\nabla \times \bu)}\right]_{j}[\bk,t]\right\}
	\end{split}
\end{equation*}
represents the energy transfer function.\\

We finally define the forcing timescale as	$\tau = \ell_{\rm f}/\sqrt{E_0}$, with 
$E_0$ denoting either the prescribed RNS energy or a suitable time-averaged  NS energy. 
Please observe that in Fourier space, the reversible viscosity defined in 
Eq.~\eqref{eq:revviscosity}, is  computed  as 

\begin{equation} \label{eq:revviscosity2}
\begin{split}
&\nur[\bu] = \epsilon_{\inj}/ \Omega, \hspace{0.2cm} \text{ with } \Omega \text{ the enstrophy}, \\
& \text{and }\epsilon_{\inj} := \Re \left\lbrace \sum_{\substack{\bk: |\bk| \le \kmax \\ i=1,2,3}} { f_i}(\bk,t) \cdot u_i^\star (\bk,t) \right\rbrace
\end{split}
\end{equation}
representing the injected power due  to the external forcing.

\subsubsection{Parameters of the Simulations}

In order to carry out a systematic investigation of the RNS system at fixed $N_c$
and fixed  forcing wave number $k_{\rm f}=\sqrt{3}$, we follow two protocols:
($\tt A$) vary $u_{0}$ so that the  runs have different $E_0$ for fixed $f_0$;
($ \tt B$) vary $f_0$ using  the same prescribed initial velocity amplitude 
$u_{0} = 1$ (in which case all the corresponding RNS runs have the same total energy). 
Our discussion is based on three sets of runs of resolution up to $128^3$, the  
details and labels of which are summarized in Table \ref{tab:simu}.
 
\begin{table}[h]
\renewcommand{\arraystretch}{1.3}
\begin{tabular}{|c|c|c|c|c|}
\hline
   Set & $N_c$ & $\kmax$&$E_0$ & $f_0$ \\
\hline
\hline
   \B & $64$  & $21.3$ & From $0.06$ to $2.2$ &$  0.13$ \\
\hline
   \Atwo & $128$&$42.6$ & From $0.06$ to $2.2$ &$  0.12$\\
\hline
   \Aone & $128$& $  42.6$& 0.125 & From $0.012$ to $0.12$\\
\hline
\end{tabular}
\caption{The three sets of runs discussed in the present  work.}
\label{tab:simu}
\end{table}

\subsection{Results}
\label{sec:results}
\begin{figure*}[tb]
\includegraphics[width=\textwidth,trim=0.5cm 1cm 1cm 1cm,clip]{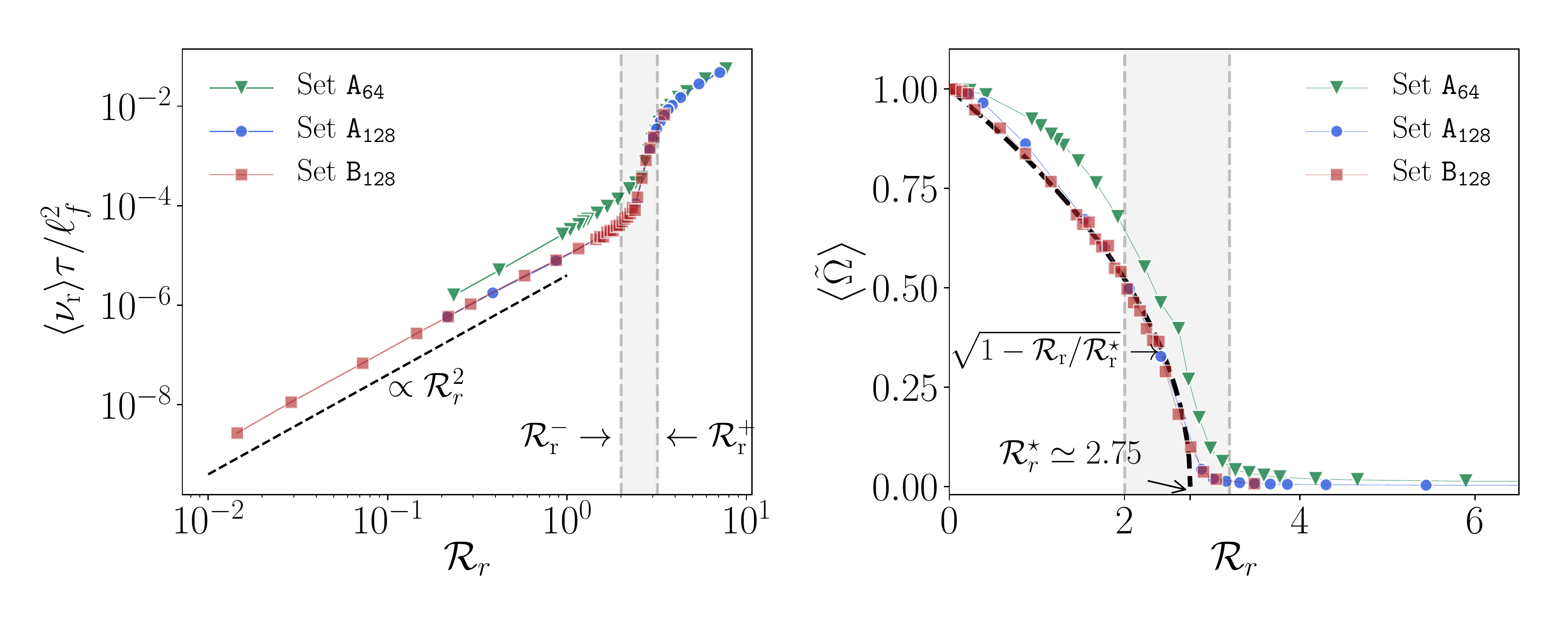}
\put(-440,10){\large{\bf(a)}}
\put(-170,10){\large{\bf(b)}}\\
\includegraphics[width=\textwidth,trim=0.5cm 1cm 1cm 0cm,clip]{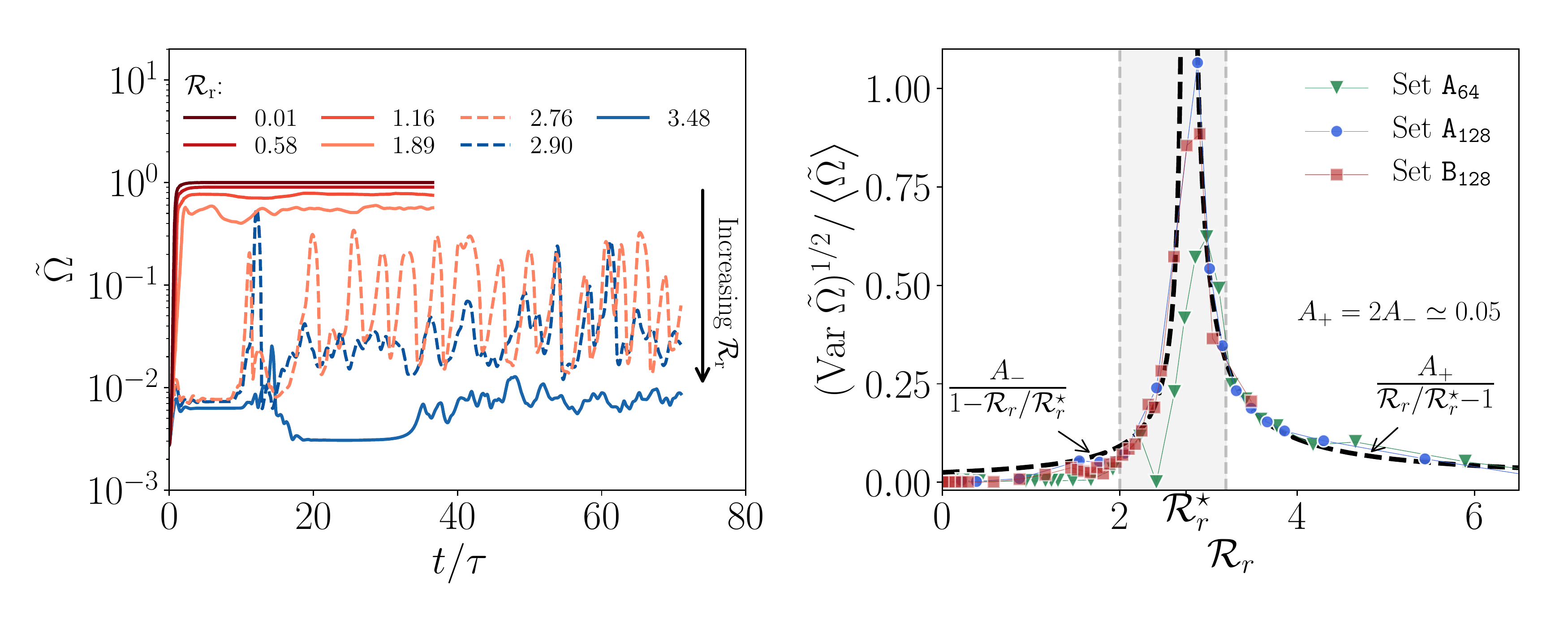}
\put(-440,10){\large{\bf(c)}}
\put(-170,10){\large{\bf(d)}}\\
\caption{{\bf Signatures of the RNS phase transition.}
	(a) Time-averaged reversible viscosity $\nu_r$ vs. $\Rr$.
	Dotted lines represent the  scaling  $\nu_r \sim  \Rr^2 $,  prescribed by  the asymptotics $\Rr \to 0$ (see text for details).
	For $\Rr > \Rr^\star$, the reversible viscosity becomes independent of the cutoff $\kmax$. This signals that the regime is of  hydrodynamic type.
	(b) Time-averaged  normalized enstrophy $\widetilde{\Omega}$ vs. $\Rr$, 
	exhibiting a smooth transition in the vicinity of  
	$\Rr^\star \simeq 2.75$.    
	$\widetilde{\Omega} \in [0,1] $  acts as an 
	\textit{order parameter}. 
	(c) Time series of the normalized enstrophy ($\widetilde{\Omega}:=\Omega/\Omega_{eq}$) for representative
	values of the  control parameter $\Rr$ from Set \Aone.
	(d) Time-variance of $\widetilde{\Omega}$ vs. $\Rr$ showing the enhancement 
	of the enstrophy fluctuations near $\Rr \sim \Rr^\star$.
	In every figure, the grey-shaded area indicates a transition range delimited by $\Rr^- \simeq 2$ and $\Rr^+ \simeq 3.2$.
}
\label{fig:globquant}
\end{figure*}
We now present our results from DNSs of the RNS system. 
We find that the time-averaged value of the enstrophy and reversible viscosity,
as a function of the control parameter $\Rr$, provides an insightful way to 
distinguish between the different statistical regimes of the RNS system.
They serve as a relevant order parameter that clearly demarcates the phase diagram
in two regions $\Rr < \Rr^- \simeq 2 $ and $\Rr > \Rr^+ \simeq 3.2 $, in accordance with
the two asymptotic states $\Rr \to 0$ and $\Rr \to \infty$.
We refer to the low-$\Rr$ regime as a ``warm'' statistical states, as they are characterized by the 
presence of a partial thermalization at small length-scales and therefore are sensitive to the 
cutoff wave number $\kmax$.
The high-$\Rr$ regime is insensitive to this non-physical cutoff, and hence we refer to these states
as being of ``hydrodynamic'' type. The range $\Rr^- \le  \Rr \le \Rr^+$ is a crossover region over which
the mean value of the order parameter smoothly decreases from a finite positive values to a nearly vanishing one.
Moreover, the vicinity of the critical point is characterized by the presence of strong fluctuations (bursts) 
in the enstrophy time series, whose origin in our understanding is linked to a non-trivial, ultra-violet multi-scale dynamics.


\subsubsection{Statistical states of the RNS system}
\label{sec:globquant}

The enstrophy $\Omega$ is particularly sensitive to the onset of a thermalization at small 
length-scales (large wave numbers). For our purposes here, we find it convenient to normalize the enstrophy as 
$\widetilde \Omega := \Omega/\Omega_{\eq}$, where $\Omega_{\eq}=3\kmax^2 E_0/5$ is the absolute equilibrium value
(see  Appendix  \ref{sec:abseq} for details).

Note that the (non-signed) reversible viscosity, defined by Eq.~\eqref{eq:revviscosity2}, is also
sensitive to the fluctuations at small length scales. Therefore, the time-averaged enstrophy and the reversible
viscosity emerge as a natural choice for the order parameter, which allows us to distinguish between the
different phases (statistical regimes) of the RNS system.
In the asymptotic limit $\Rr \to 0$, i.e. in case of full thermalization, $\av{\nur} \to 0$ 
and $ \langle \tOm \rangle \to 1$, while the opposite limit $\Rr \to \infty$ (over-damped regime)
corresponds to  $\av\nur \to \infty$ and $ \langle \tOm \rangle\to  0$.

Figures~\ref{fig:globquant} (a)-(b) show the values of $\av{\nur} $ and $ \langle \tOm \rangle$
as function of the control parameter $\Rr$, on which the crudely depicted ``crossover region'' of \figref{fig:SketchI} between the  small-$\Rr$ \emph{thermalized} and the large-$\Rr$ \emph{hydrodynamical}
can be identified as the critical region in the close vicinity of $\Rr^\star = 2.75$.

This information can be refined by monitoring the dynamical behavior of the order parameters, 
rather than their time-averaged values.
To simplify the discussion, we choose to comment only on the fluctuations of the normalized enstrophy, 
which we show in~\figref{fig:globquant}(c) and  (d). Specifically, \figref{fig:globquant}~(c) displays 
the dynamical evolution $\tOm$ for representative values of $\Rr$ corresponding to the RNS Set \Aone, 
while \figref{fig:globquant}~(d) shows the relative amplitude of the fluctuations 
(standard-deviation of $\tOm$ relative to $\langle \tOm \rangle$) for all three sets of runs \B, \Atwo~ 
and \Aone~as a function of the control parameter $\Rr$.
The combined insight from both the averaged and dynamical behavior of $\tOm$ leads us to identify three sub-ranges of the control 
parameter $\Rr$ that correspond to three different statistical regimes of the RNS system that we describe below.


\paragraph{The hydrodynamic range:} $\Rr > \Rr^+ \simeq 3.2$.\\
In this range, the time-averaged reversible viscosities $\av \nur$ reach finite positive values. 
The data collapse observed in \figref{fig:globquant} (a) suggests that these values are described by an  increasing function 
of $\Rr$, independent of both the chosen protocol  and the cutoff scale $\kmax$. 
This feature is compatible with the normalized enstrophy $\langle \tOm \rangle$ being nearly vanishing and behaving as a decreasing 
function of $\Rr$. Note that $\langle \tOm \rangle$ has some dependence on $\kmax$,
this  is evident from the observation that its profile for set \B\, in \figref{fig:globquant} (b)  lies above those for sets \Atwo\, and \Aone.

Furthermore, the enstrophy time-series indicate  that in this range of $\Rr$,  typical statistically steady states have  
low-amplitude enstrophy fluctuations. 
Empirical fits shown in \figref{fig:globquant} (d) reveal that the normalized temporal standard-deviations of the enstrophies 
are reasonably well described by $[\text{Var}( \tOm )]^{1/2}\simeq A_+ \langle \tOm \rangle/(\Rr/\Rr^\star-1)$, 
with $A_+ = 0.05$ and $\Rr^\star = 2.75$, independent of $\kmax$.\\

\paragraph{The warm range:} $\Rr < \Rr^- \simeq 2$.\\
In this range of $\Rr$, the order parameters show some dependence on $\kmax$, as visible in \figref{fig:globquant} (a) and (b).
For the reversible viscosity, this can be  accounted for by using the Kubo dissipation theorem to estimate  
$\av{\epsilon_\inj} \sim f_0^2 \tau_{\eq}$ in the limit $\Rr \to 0$.
The timescale $\tau_\eq \sim \lf E_0^{-1/2} $ is the \emph{equilibrium} velocity correlation time at forcing length-scale, and is here prescribed by the 
statistics of the fully thermalized state of the truncated Euler flows \cite{kubo1957statistical,thalabard2017optimal}.
Combining this estimate with the definitions \eqref{eq:dimlessnoRNS} and \eqref{eq:revviscosity2},  and  using Eq.\eqref{Eq:OmegaEQabs} found in Appendix \ref{sec:abseq}, one obtains 
\begin{equation}
	 \label{eq:warmRNSnu}
	\av \nur 	 \sim \dfrac{\Rr^2 E_0^{3/2}}{\Omega_\eq \lf} \propto \dfrac{\Rr^2}{k_{\max}^2} \,\,\text{as}\,\, \Rr \to 0.
\end{equation}
This  asymptotics indeed accounts for the scaling behaviors observed in our numerics, which in fact is present in the entire warm range of $\Rr$.

For the normalized enstrophy,  the time-average $\langle \tOm \rangle$ observed for the higher resolved sets of runs  
prove to be very accurately fitted by  the square-root profile $\langle  \tOm \rangle = (1-\Rr/\Rr^\star)^{1/2}$. 
The representative time-series of \figref{fig:globquant} (c) indicate that warm dynamics quickly reach steady states characterized by vanishing levels of fluctuations for the enstrophy. As a function of $\Rr$, these  are  fairly well described by the fit $[\text{Var}( \tOm )]^{1/2}\simeq A_- \langle \tOm \rangle/(1-\Rr/\Rr^\star)$, with $A_- = 0.025$.\\

At the present stage,  the specific shapes of the fitting profiles should be considered as  mere observations. 
Clearly, these  fitting laws indicate a critical behavior, which is amenable to a mean-field treatment and is suggestive of 
a potential continuous phase transition. 
Yet,  we  postpone any further informed comments on the mean-field treatment up until \S~\ref{sec:landau}, 
where similar behaviors  will  again appear but in a somewhat simplified setting, hence it is easier to gain insight. \\

\paragraph{The transition range: }$\Rr-<\Rr < \Rr^+$.\\
Within this narrow range of $\Rr$, the order parameters sharply but \emph{smoothly} transit between their  warm  and 
hydrodynamic behaviors:  This  precisely corresponds to the  crossover region anticipated in  \figref{fig:SketchI}. 
Let us observe that the critical value $\Rr^\star\simeq 2.75$ previously obtained as a fitting parameter lies in this range.
In fact, the mixed phase is essentially identified from the dynamical behavior of the  enstrophy that becomes bursty,
 characterized by the appearance of successive peaks (\emph{cf.} \figref{fig:globquant}~(c)). 
 This behavior is found to persist up to the maximal integration time that we considered. 
 The bursty behavior implies that the enstrophy fluctuations get drastically enhanced with respect to their the mean enstrophy 
 values   when approaching $\Rr^\star$ from either the warm or the hydrodynamic side, as is evident form  \figref{fig:globquant} (d). 
 At  $\Rr =\Rr^\star$ for example,  we observe $[\textrm{Var}(\widetilde{\Omega})]^{1/2}/\langle\widetilde{\Omega}\rangle\sim 1$ 
for all three sets of runs, meaning that the temporal fluctuations are of the order of the time-averages. 
This behavior indicates finite-size effects and an associated potential continuous phase transition that occurs in 
the limit of infinite resolution, i.e.  $\kmax \to \infty$.

\old
\subsubsection{Spectral signatures of the RNS states}

Here we document the RNS energy spectra and fluxes  observed in the different statistical regimes,
with an aim to further characterize the phase diagram of the RNS system, without associating them
with their NS counterpart or the equivalence conjecture, at least for now.
Let us recall that the RNS dissipative term relies on an intrinsic direct dynamical coupling 
with the forcing length-scale. It  is therefore highly non-local in Fourier space, in sharp 
contrast with the standard NS viscous damping, which is local in $k$-space.  
\emph{A priori}, the spectral signatures of the different RNS states are not obvious; 
it is unclear whether we should at all expect the RNS system to even mimic the standard NS
\emph{phenomenology} for prescribed ranges of $\Rr$. \\

\paragraph{Warm spectra \emph{vs.} hydrodynamic spectra.}\hspace{0.1cm}\\
\label{sec:RNSWHspectra}
\begin{figure*}[htb]
\includegraphics[width=\textwidth,trim=0.5cm 1cm 1cm 0cm,clip]{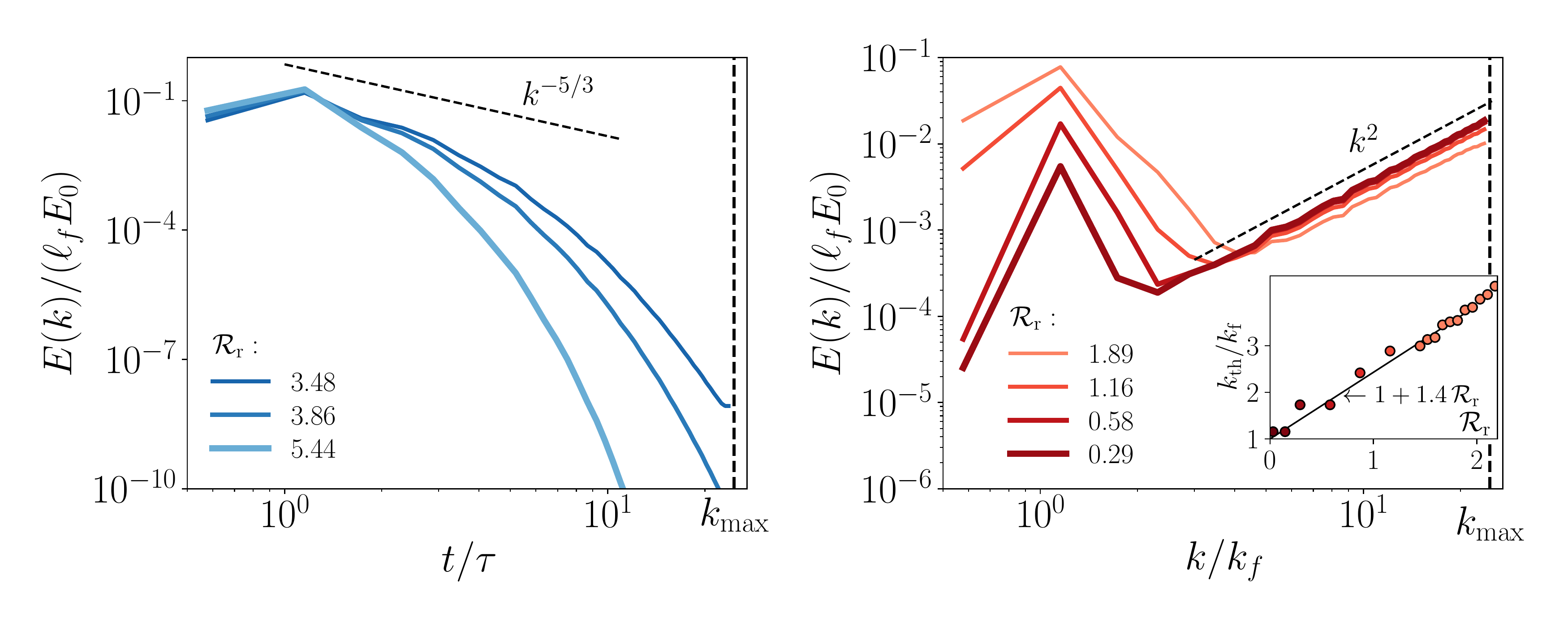}
\put(-450,10){\large{\bf(a)}}
\put(-190,10){\large{\bf(b)}}
\caption{
{\bf Warm spectra \emph{vs.} hydrodynamic spectra.}
 Main panels show the suitably normalized time-averaged  energy spectra  observed in the RNS runs at $N_c=128$ (from Set \Aone\, and for $\Rr>3.48$ from Set \Atwo\,) for representative values of: 
	(a) $ \Rr > \Rr^+$  corresponding to the hydrodynamic regime;
	(b) $ \Rr < \Rr^-$ corresponding to the warm regime. 
	In the latter case, we observe  an increasing range of wave numbers $k>k_\th$ over which the spectra have Gibbsian scaling $E(k) \propto k^2$. $k_\th$ is the wave number at which $E(k)$ is minimal.  
	The inset  reveals a linear dependence of  $k_\th$ on $\Rr$, and suggests  $ k_\th \to \kf \simeq 1$ as $\Rr \to 0$.
}
\label{fig:RNSspecWH}
\end{figure*}
The analysis of \S~\ref{sec:globquant} revealed that  within the warm and the hydrodynamic phases, RNS dynamics has  non-equilibrium 
steady states characterized by very low enstrophy fluctuations. In  both of these phases, it is therefore natural to focus on time-averaged 
quantities. We define the  (stationary) energy spectrum as the time-average $E(k) := \av{E(k,t)}$, where the bracket indicates an average 
over the total duration of the simulations.  
As shown in \figref{fig:RNSspecWH}, both the warm and the hydrodynamical phase have distinct spectral signatures, 
which naturally links with the behavior of the order parameter studied in \S~\ref{sec:globquant}.\\

In the hydrodynamical phase, $\Rr >\Rr^+$, the energy spectra have compact support in $k$-space, as shown in \figref{fig:RNSspecWH} (a).
For $\Rr \gg R_+$, the supports is narrow, the spectra being contained within a small $k$-range around the forcing wave number. 
This means that the effective scale-by-scale damping mechanism generated from the reversible  viscosity is large and 
dominates over the non-linear transfer, somewhat akin to a laminar regime.
As $\Rr$ decreases down to $\Rr^+$, energy  spreads towards the higher 
wave numbers $k > \kf$.  The system is then in a non-trivial non-equilibrium steady-state, with non-zero flux of energy, and 
multi-scale statistics  being essentially independent of $\kmax$: This could be taken as a heuristic 
definition of a turbulent state \cite{Falkovich2008Introduction}!
From this qualitative point of view, the RNS statistics observed at $\Rr = 3.48 \gtrsim \Rr^+$ do indeed describe a turbulent motion.

At this stage, the rather modest resolution of our numerics compared to current state-of-the-art  NS simulations precludes us from  
drawing any conclusion as to whether  higher-resolved RNS simulations would indeed produce \emph{fully developed turbulent statistics}, 
\emph{e.g.}, akin to those found in  numerical and experimental datasets related to  extreme regimes of fluid motion
\cite{yeung2015extreme,saw2016experimental}. This issue is related to the equivalence conjecture and we will discuss it in 
 details in \S~\ref{sec:TurbulentLimit}, in connection with our discussion on turbulent limits. \\

In the warm phase,  here identified as $\Rr<\Rr- \simeq 2$, the spectra are contaminated by the finite cutoff, as shown in \figref{fig:RNSspecWH} (b). 
Specifically, they resemble some of the transients commonly observed in numerical simulations of the  truncated Euler dynamics \cite{cichowlas2005effective,krstulovic2008two,krstulovic2009cascades}, in the sense that a seemingly  infra-red traditional hydrodynamic scaling  at small $k$ coexists with a near  equilibrium ultra-violet power law scaling, that is $E(k) \sim k^\alpha$, where  $\alpha$   progressively increases towards the Gibbs exponent $\alpha=2$ as $\Rr$ decreases towards 0.
 The separation between the two regions is  identified in terms of a thermal wave number $k_\th$, defined as the local minimum of the energy profile. The inset of \figref{fig:RNSspecWH}~(b) shows that as $\Rr\to 0$, $k_{\th}$ decreases linearly towards a value close to the forcing scale $k_\th = \sqrt 3$, \emph{e.g.} close to  the smallest wavenumber $k_0=1$. This is compatible with the fact that at $\Rr = 0$, the RNS steady state in fact corresponds to a fully thermalized \emph{equilibrium} state of the truncated Euler equations, as explained in \S~\ref{ssec:Rr}.
Naturally,  the approximate ultra-violet thermalization at $k>k_\th$ accounts for the fact that  the warm phase has a non-vanishing order parameter $\langle \tOm\rangle$, as indeed observed in \figref{fig:globquant}, and explained with more technical details in Appendix \ref{sec:abseq}.\\

\paragraph{Energy spectra in the transition range.}\hspace{0.1cm}\\
\label{sec:RNStransspectra}
\begin{figure*}[htb]
\includegraphics[width=\textwidth,trim=0.5cm 1cm 1cm 0cm,clip]{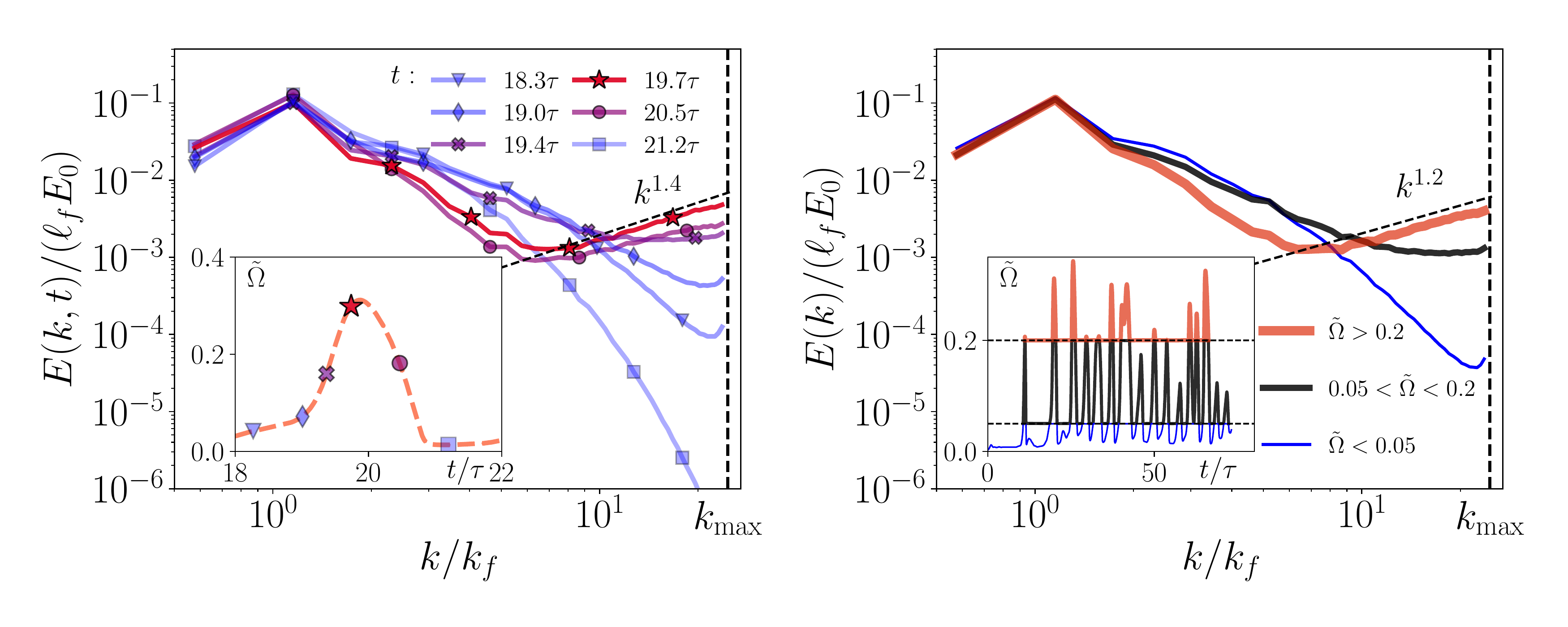}
\put(-450,10){\large{\bf(a)}}
\put(-190,10){\large{\bf(b)}}
\caption{
{\bf Spectral multi-stability in the  transition region,  $\Rr^-<\Rr < \Rr^+$.} 
	Panel (a) shows  duly normalized \emph{instantaneous} energy spectra from the RNS run  $\Rr = 2.76$ of Set \Aone, at selected times, so as to  trace one of the enstrophy burst shown in \figref{fig:globquant} (c).
	 Inset shows the zoom of the enstrophy burst in the time series. 
	Panel (b) displays  time-averages of the energy spectra, conditioned   on specific  magnitudes of the normalized enstrophy, as specified by the inset.
}
\label{fig:fluctespec}
\end{figure*}
In the transition region, as evidenced by the violent fluctuations observed in the enstrophy time series, it is unclear whether the system genuinely reaches a steady state. It proves therefore more instructive  to comment on the dynamics of the instantaneous energy spectra $E(k,t)$, rather than on their  time-averaged values.
In fact, the peaks observed in \figref{fig:globquant} (c) clearly relates to oscillations of the ultra-violet behavior near $\kmax$.
We illustrate this by using the RNS run  of set \Aone, corresponding to  $\Rr =2.76$, a value close to the identified critical point $\Rr^\star =2.75$ at which the fluctuations are the most enhanced.
\figref{fig:fluctespec}(a) reports the dynamical evolution of the energy spectra $E(k,t)$ on a short time interval $18 \tau <t <22\tau$, over which the normalized enstrophy  abruptly varies from $\tOm  \simeq 0.03$  at $t \simeq 18.3\tau $ to $\tOm \simeq 0.3$  at $t \simeq 19.7 \tau$ back to $\tOm \simeq 0.01$ at $t = 21.2 \tau$.
Over this time interval, the infra-red energy spectra near the forcing length-scale remains essentially unchanged, but the ultra-violet profile drastically varies. It transits between being  exponentially damped and being algebraic, with  time-dependent scaling $E(k,t) \propto k^{\alpha(t)}$, over a scaling range whose size increases with the exponent $\alpha$.

As the enstrophy increases in time towards its peak value, the exponent $\alpha(t)$ itself switches from negative to positive values, 
and the scaling-range develops on a gradually increasing range. Let us observe,  that the maximum value reached by the scaling 
exponent is $1.4$ and not $2$, as would be expected if the system was partially thermalized.  

We know that  the enstrophy is  particularly sensitive to the dynamical evolution of small length-scales. 
This is further illustrated in \figref{fig:fluctespec}, where  time-averages of the energy spectra conditioned on prescribed enstrophy values are indeed observed to yield very different ultra-violet scaling ranges.
For example, conditioning on $\tOm< 0.05$ yields a close-to-hydrodynamic type spectrum,  while conditioning  on the highest values $\tOm>0.2$ produce  ultra-violet scaling reminiscent of a warm one. 
Again, a closer inspection reveals that the relevant scaling exponent is  only  $1.2 $ and not $2$.

This implies that while a finite $\kmax$ indeed produces non-zero values for the order parameter $\langle \tOm \rangle$, 
the latter should vanish in the limit $\kmax \to \infty$. This naturally hints that the crossover range should disappear in this limit.
 This would imply $\Rr^- =\Rr^+=\Rr^\star$ asymptotically, and strongly suggests that the transition between the warm states and the hydrodynamics states becomes a genuine continuous phase transition in the limit $\kmax \to \infty$.\\

\paragraph{Energy fluxes.}\hspace{0.1cm}\\
\begin{figure*}[htb]
\includegraphics[width=\textwidth,trim=0.5cm 1cm 1cm 0cm,clip]{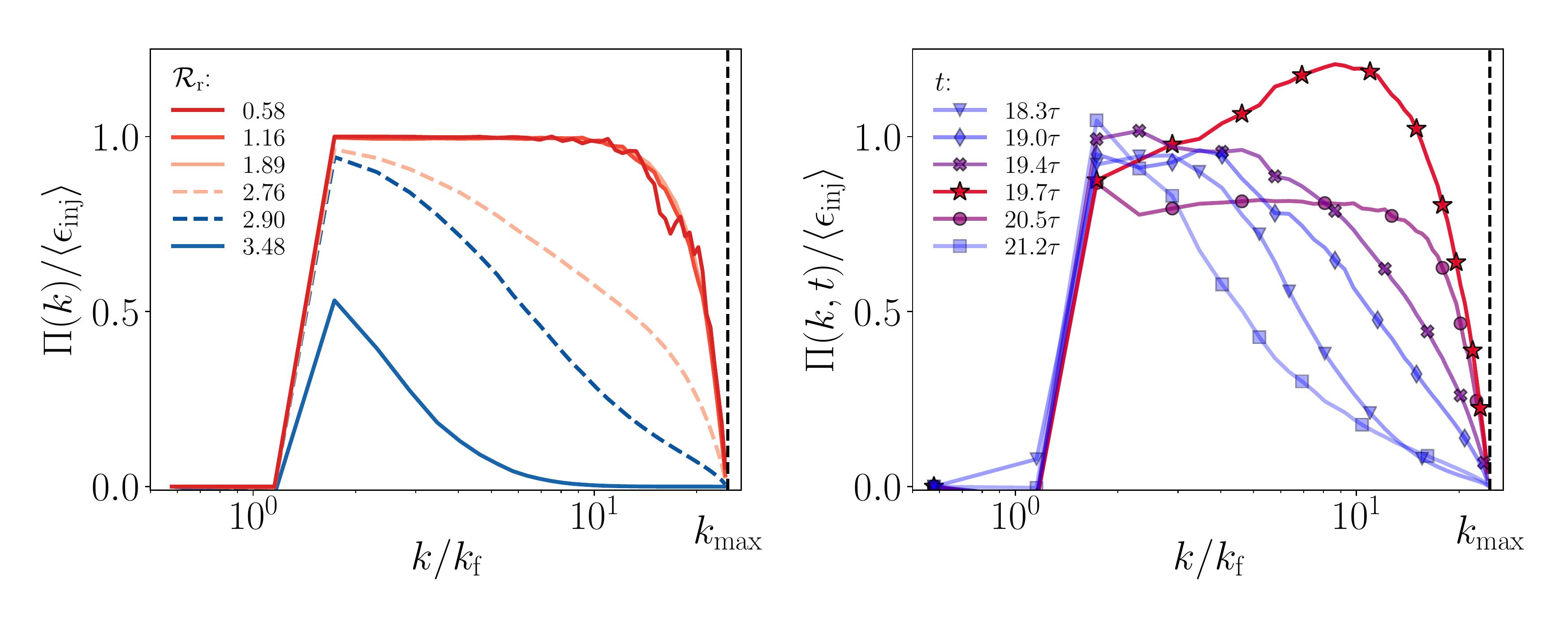}
\put(-460,10){\large{\bf(a)}}
\put(-200,10){\large{\bf(b)}}
\caption{
{\bf RNS Energy fluxes.} Panels show the $k$-space profile of the  energy fluxes $\Pi(k)$  normalized by the time-averaged injected power $\av \epsilon_\inj$, computed from Set \Aone\,: 
(a) averaged over time for  values of $\Rr$ representing different RNS regimes;
(b) for the $\Rr=2.76$ run within the crossover region, at the same specific  times  considered in Fig. \ref{fig:fluctespec} (a) and tracing the enstrophy burst shown in the corresponding inset.
}
\label{fig:avgfluxdifff0}
\end{figure*}

In order to conclude our overview of the RNS states, let us briefly comment on the RNS energy fluxes, whose profiles 
for  set \Aone\, are shown in \figref{fig:avgfluxdifff0}, wherein they have been normalized by the time-averaged injected 
power $\langle\epsilon_\inj\rangle$. The transition from the hydrodynamic to the warm regime is reflected by  
the $k$-space profiles of the time-averaged fluxes   $\Pi(k):=\av{\Pi(k,t)}$, as shown in \figref{fig:avgfluxdifff0}~(a).\\

The  flux profiles observed within the hydrodynamic range clearly mirror  the energy spectra observations of
\S~\ref{sec:RNSWHspectra}. For large $\Rr > \Rr^+$,  fluxes are indeed non-zero only in a small range of wave numbers $k \gtrsim \kf $. As $\Rr$ decreases down to $\Rr^+$, the spectral extension of the fluxes increases. For example, at $\Rr \approx 3.48$, where the energy spectra suggest that  RNS dynamics is  multi-scale, we 
observe an energy flux that is significant up to $k \approx 10\,\kf$: This is in qualitative agreement with standard  NS phenomenology at low Reynolds number. In particular, it is not evident from the fluxes that the reversible dissipation is defined as a non-local operator in $k$-space.\\

Within  the warm range $\Rr <\Rr^-$, the  fluxes follow a  universal profile, that is seemingly independent of the specific value of $\Rr$, as indicated by the data collapse observed for  $ 0.58 \lesssim \Rr \lesssim 1.89$.
This is a clear  signature of the warm statistical regime, but is counter-intuitive.

On the one hand, except for their abrupt decay to zero at the ultra-violet end, the fluxes are  mostly constant over 
the entire $k$-range  above the forcing scale, \emph{e.g.} here   $2\kf \lesssim k \lesssim20 \kf $:  This  signals an 
out-of-equilibrium state, that should imply   Kolmogorov-scaling  for the spectra.  
On the other hand, the corresponding energy spectra do not show this Kolmogorov spectra. Instead,  \figref{fig:RNSspecWH}~(b) indicates
close-to-equilibrium statistics, with a ``distance'' towards full thermalization monitored by the scale  $k_\th$ becoming arbitrarily close to $1$ as $\Rr \to 0$. This gradual convergence towards the equipartition state is not  reflected by the flux profiles. 

A qualitative explanation could be that the energy flux is an integrated quantity (see \S~\ref{sec:conventions}). Hence, if the  
range  $k >k_\th$ is indeed nearly thermalized as prescribed by the equipartition statistics, then they  do not contribute to 
the flux.  An impatient and puzzled  reader can however jump to  the discussion of Eq.~\eqref{eq:warmprofile} at the end 
of \S~\ref{sec:RLqualitative} to find that this kind of profile is in fact fully compatible with near-equilibrium and partially thermalized statistics.

\old
As an aside, let us point out that in the ``warmest range'' $\Rr \lesssim 0.5$, not shown here,  $\Pi(k)$ begins to fluctuate wildly  from the ultra-violet end.  The amplitude of the normalized fluctuation grows with decreasing 
$\Rr$ and destroys the plateau behavior. The amplitude of the non-normalized $\Pi(k)$ however correctly goes to zero and this is compatible with the  truncated Euler limit with $E(k)\sim k^2$ across the full $k$-range.\\

Within the transition range $\Rr^-<\Rr <\Rr^+$, \figref{fig:avgfluxdifff0} indicates that the large enstrophy fluctuations reflect the fact that the instantaneous fluxes $\Pi(k,t)$ oscillate in time between a  narrow-band hydrodynamic-type profile (at $t=21.2 \tau$ for example) and a multitude 
of full-band profiles, which for instance include the constant profile at $t=20.5 \tau$ or the non-monotonic  bumpy profile peaked at $k \simeq 10 \kf$ in the vicinity of  $\kmax$ at $t=19.7 \tau$, and corresponds to the local maximum monitored in  the enstrophy time series.

\section{Insights from a reversible Leith-type toy model}
\label{sec:Leith}
Our numerical analysis so far shows  that the RNS system undergoes a continuous phase transition at 
the critical point $\Rr^\star$, whereby steady RNS solutions transit from  being
\emph{hydrodynamic} to being \emph{warm}, in the sense that their ultra-violet statistical 
features become affected by truncation scale and eventually  thermalize. 
The smoothness of the transition is however a necessary consequence of  our RNS runs having a finite resolution. 
Even though the behavior of the order-parameters (see~\figref{fig:globquant}) appears to be consistent with  the phenomenology 
a second-order transition occurring at $\Rr^\star \simeq 2.75$, the numerical evidence is only suggestive.
Our runs have finite resolutions, and this in principle precludes true divergence of any first derivative of the control parameter  
at the candidate critical point.
For a similar reason, while we argued that  the RNS equations at $\Rr \gtrsim \mathcal \Rr^\star$  produce multi-scale steady states 
that fit the heuristic  definition of ``turbulence''; we are well aware that such  a statement is  only qualitative,  
due to the modest resolutions of our RNS runs. 
Consequently,  it cannot provide  a  firm assessment regarding the validity of the equivalence  conjecture. 
This is the reason why no quantitative comparison with NS runs have been commented on so far.\\

Even though higher resolutions simulations are desirable, nevertheless, further insights about the nature of the transition can
 be obtained at a smaller numerical cost from a simplified non-linear diffusion spectral 
model of turbulence, namely a modified Leith model of turbulent cascade ~\cite{leith1967diffusion}, 
which mimics the statistical properties of the RNS system. 
This model is easier to analyze as its steady solutions can determined semi-analytically
~\cite{connaughton2004warm,connaughton2011mixed,thalabard2015anomalous,grebenev2016SSLM}.
Note that the terminology ``warm solutions'' used in the present paper is borrowed from the concept
of ``warm cascades'' introduced in Ref.~\cite{connaughton2004warm}, which is a stationary solution of
the inviscid Leith model that exhibits simultaneously the Kolmogorov infra-red scaling and thermalized Gibbsian ultra-violet statistics.

Our analysis of the reversible Leith model suggests that its steady solutions indeed undergo a second-order phase transition, which
separates the hydrodynamic scaling from the warm solutions. The transition is controlled by a parameter $\mathcal R_L$, similar to
the control parameter $\Rr$ of the RNS system, and the phase diagram at finite $\kmax$ is qualitatively similar to the one inferred from 
the RNS simulations. The parameter $1/\kmax$ plays the role of  a symmetry breaking parameter, 
which is reminiscent of the magnetic field in the Ising model.
In the limit $\kmax \to \infty$, this system undergoes a genuine continuous phase transition, at which a suitably defined susceptibility 
diverges. Moreover, we find that the statistical features of this phase transition can be captured by constructing a mean-field Landau 
free energy. We argue that this picture extends to the RNS system and has practical implications for the equivalence conjecture.

\subsection{Description of the Reversible Leith model}
The inviscid  Leith model ~\cite{connaughton2004warm} consists in approximating  dynamics of the energy spectrum in  $k$-space using  a well-chosen  second-order non-linear diffusive operator. We here combine this non-linear evolution for the energy profile $E(k,t)$ with a thermostat.
The Reversible Leith (RL) dynamics is then simply described by  
\begin{equation}
\begin{split}
\label{eq:Leithmodel}
&\frac{\partial E(k,t)}{\partial t} = -\frac{\partial \pi(k,t)}{\partial k} - \nu_Lk^2 E(k,t),\\
& \text{where } \hspace{0.2cm } \pi(k,t) = -C k^{11/2} E^{1/2}(k,t) \frac{\partial}{\partial k} \Bigl[\frac{E(k,t)}{k^{2}} \Bigr]
\end{split}
\end{equation}
represents an energy flux and $C$ is a dimensional constant that can be set to $1$ for the present purpose. 
The wave numbers $k$ range from prescribed $k_0 $ to the truncation wave number $ \kmax$.
In analogy with the RNS system, we interpret the parameter $\nu_L$ as  a reversible viscosity that guarantees the conservation of the total energy
\begin{equation}
	\int_{k_0}^{ \kmax} E(k,t) \,d k = E_0 \hspace{0.2cm}\text{(prescribed).}
\end{equation}

We seek to characterize the non-equilibrium steady energy spectra $E(k)$ and the associated  flux $\pi(k)$, generated by the RL dynamics~(\ref{eq:Leithmodel}), when the 
following fluxes are prescribed  at the boundaries~:
\begin{equation}
\label{eq:LeithBC}
\pi(k_0) = \epsilon_0 \;\; 
\text{and} \;\;  \pi(k_{max}) = 0.
\end{equation}

Combining the stationarity condition 
\begin{equation}
\label{eq:LMconstEsteady}
-\frac{\partial \pi(k)}{\partial k} = \nu_L k^2 E(k)
\end{equation}
with the boundary flux conditions Eq.~\eqref{eq:LeithBC}, the reversible viscosity can be 
explicitly tied to the stationary energy profile $E(k)$ as 
\begin{equation}
\label{eq:LeithViscosity}
	\nu_L= \dfrac{\epsilon_0}{\int_{k_0}^{\kmax} k^2 E(k)\,dk}.
\end{equation}

The independent parameters that control the behavior of the steady energy profile  
are $k_0$, $\kmax$, $E_0$,  together with the infra-red boundary  flux $\epsilon_0$. 
Letting $\kmax \to \infty$ and $\nu_L \to 0$ allows the possibility of non-vanishing constant flux solutions, characterized
by the Kolmogorov scaling $E(k) \sim k^{-5/3}$. 
Such Kolmogorov solutions have finite capacity spectra, namely 
$\int_{k_0}^{+\infty}  E(k) \,dk <\infty$. In other words, the value of the total energy is  independent of $\kmax$ when $\kmax \to \infty$; 
therefore, it is  natural to  define a 
dimensionless number independent of $\kmax$, as 
\begin{equation}
	\label{eq:RLparam}
	\RL = \epsilon_0^{2/3} \ell_0^{2/3} {E_0}^{-1}, \text{  with  } \ell_0 = 2 \pi/k_0.
\end{equation}
The factor $2\pi$ entering the definition of the small scale $\ell_0$ is purely cosmetic.
Although it is defined in terms of a flux rather than in terms of a forcing intensity, the dimensionless number $\RL$ is the Leith analogue of our previously defined  $\Rr$ for the RNS system.
It is the ratio of the injected energy at scale $\ell_0$ and the total energy present in the system.

\subsection{Construction of RL steady solutions}

\subsubsection{Grebenev parametrization}

In order to construct  steady solutions for the RL dynamics without resorting to direct numerical simulations of Eq.\eqref{eq:Leithmodel}, we resort to the general strategy described in \cite{grebenev2016SSLM}.  The idea is to introduce a suitable parametrization (hereafter referred to  as the ``Grebenev parametrization'') of the energy  profile that transforms  the stationarity condition Eq.~\eqref{eq:LMconstEsteady} into an autonomous bi-dimensional dynamical system.\\

\begin{figure}
	\centering
	\includegraphics[width=\columnwidth,trim=0.5cm 1cm 1cm 0cm,clip]{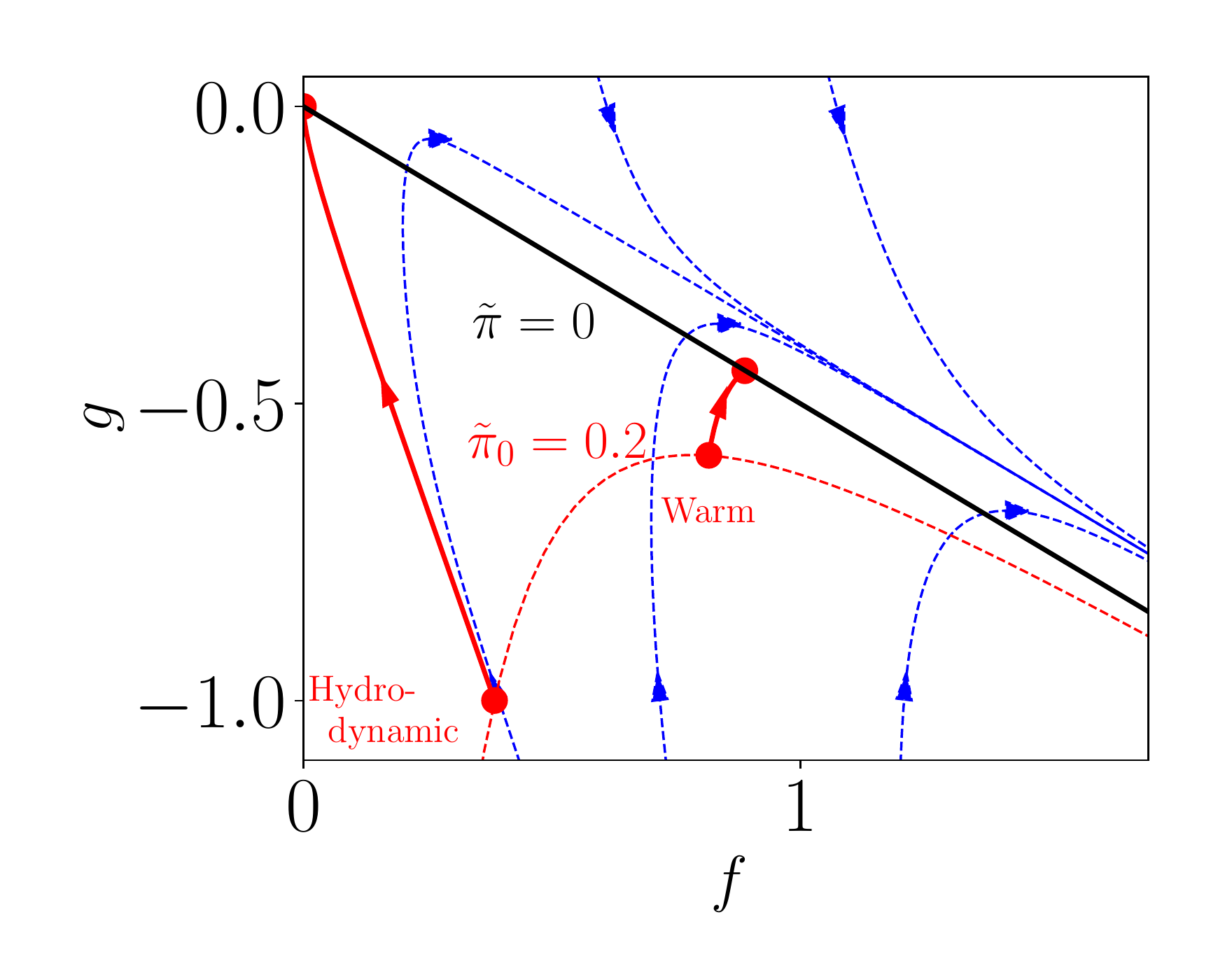}
	\caption{Phase portrait of the dynamical system Eq.~\eqref{eq:grebode} obtained from the Grebenev parametrization of the Leith stationary profiles. Blue lines represent orbits and solid red lines are two examples of parameterizing trajectories with the same  non-normalized infra-red flux $\tilde \pi_0$.  A warm solution with vanishing end-point energy flux  follows a typical orbit up until the black line, while a scaling solution follows the single orbit that ends at $(0,0)$. Note that the two represented solutions have different $\kmax$.}
	\label{fig:phaseportrait}.
\end{figure}

The specific form of the Grebenev parametrization is not particularly intuitive, but proves to be highly efficient.  
It consists in describing the steady energy profile using the  change of variables  $k,E(k),E^\prime(k) \to \tau,f(\tau),g(\tau)$
defined through
\begin{equation}
\begin{split}
	\tau := &\nu_L^{1/2} \int_{k_0}^{k} {d \kappa}\, {(\kappa E(\kappa))^{-1/2}},\\
	f(\tau) := \left({ E(k) \nu_L^{-1} k^{-1}}\right)^{1/2}&, \hspace{0.2cm} \text{and}\hspace{0.2cm}
	g(\tau) := f^\prime/f-f.
\label{eq:grebparam}
\end{split}
\end{equation}
In terms of these variables, the stationary condition \eqref{eq:LMconstEsteady} transforms into the dynamical system
\begin{equation}
\begin{split}
f^\prime(\tau) &= f(f+g),\\
g^\prime(\tau) & = -2(f+g)^2 - \dfrac{7}{2} f(f+g)+ 2 f^2+ \dfrac{1}{2} f.\\
\end{split}
\label{eq:grebode}
\end{equation}
This system admits a stable fixed point $(0,0)$ and its  phase portrait is shown in \figref{fig:phaseportrait}.

\old
\subsubsection{Practical use of the  Grebenev parametrization}
For a practical use of the Grebenev parametrization \eqref{eq:grebparam}, it is convenient to work with the non-normalized energy spectra and  fluxes, respectively defined as
\begin{equation*}
\begin{split}
&\tilde E(k) := E(k)/\nu_L^2 =  k f^2(\tau), \;\; \\
\text{and} \;\; &\tilde \pi(k) := -\pi(k)/\nu_L^3 =  k^{4} f^{2}(f+2g).
\end{split}
\end{equation*}
For given $k_0$ and $\kmax$, we can then obtain  RL steady solutions  by integrating the system Eq.~\eqref{eq:grebode} from $\tau=0$ with initial conditions $f_0,g_0$ until time $\tau_{\rm max}$, implicitly defined from Eq. \eqref{eq:grebparam} as $\tau_{\rm max} =\int_{k_0}^{\kmax} dk/(k f(\tau))$.
To construct the  admissible solutions that satisfy the  boundary conditions of type Eq.~\eqref{eq:LeithBC}, we proceed in the following manner:%
\begin{enumerate}
	\item Pick an initial value $\tilde \pi_0$ for the non-normalized flux at point $k_0$.
	Initial admissible $(f_0,g_0)$ are then such that  $\tilde \pi_0= k_0^4 f_0^2(f_0+2g_0)$.
	\item Find $(f_0,g_0)$, such that $f (\tau_{\rm max}|f_0,g_0)+ 2 g(\tau_{max}|f_0,g_0 ) = 0 $. This ensures  that $\pi(\kmax) = 0$ (up to some prescribed threshold).
	\item Compute $\nu_L= \left( \int_{k_0}^{\kmax} d k \tilde E(k)\right)^{-1/2}$.
	\item Deduce $\epsilon_0 = \nu_L^3 \tilde \pi_0$.
\end{enumerate}
The resulting solution is a steady-solution for the RL dynamics, with infra-red flux $\epsilon_0$. Examples of trajectories in the $(f,g)$-plane that parametrize either a hydrodynamic solution or a warm solution  are represented in Fig.~\ref{fig:phaseportrait}.

\subsection{Transition between Warm and Hydrodynamical steady states}
\subsubsection{Qualitative overview}
\label{sec:RLqualitative}
Using the Grebenev parametrization, we generate the RL steady energy and flux profiles for fixed $k_0=10^{-2}$, and various $\kmax$ ranging from $5 k_0$ to $1000 k_0$. For each pair $(k_0,\kmax)$, we typically vary the non-normalized infra-red flux  $\tilde \pi_0$ from $10^{-10}$ to $10^{10}$. The total energy $E_0$ is set to unity.
We observe that  at fixed $k_0$ and $\kmax$,   the steady RL solutions  are uniquely 
determined by the value of the infrared-boundary flux $\epsilon_0$. The corresponding values of 
the reversible viscosity are  then uniquely determined.\\
\begin{figure*}
	\includegraphics[width=1\textwidth,trim=0.5cm 1cm 1cm 0cm,clip]{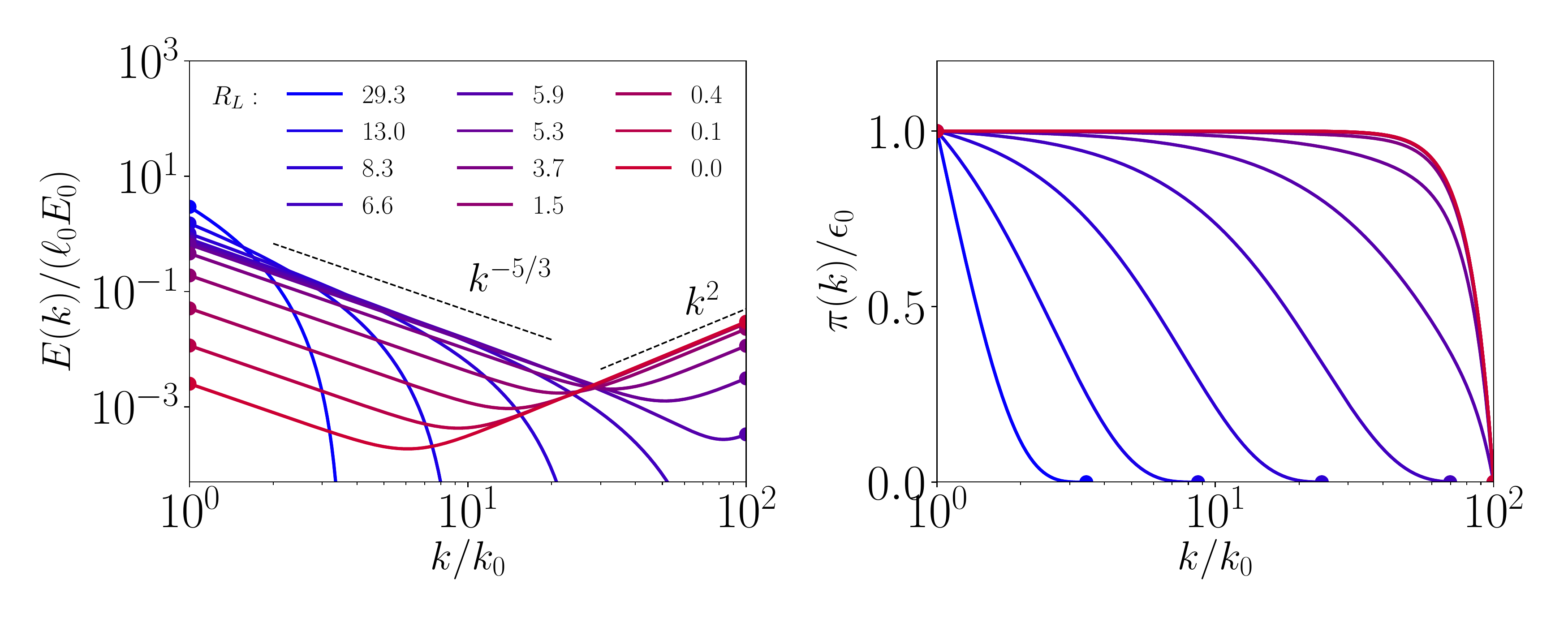}
	\put(-450,10){\large{\bf(a)}}
	\put(-190,10){\large{\bf(b)}}
	\\
	\includegraphics[width=1\textwidth,trim=0.5cm 1cm 1cm 0cm,clip]{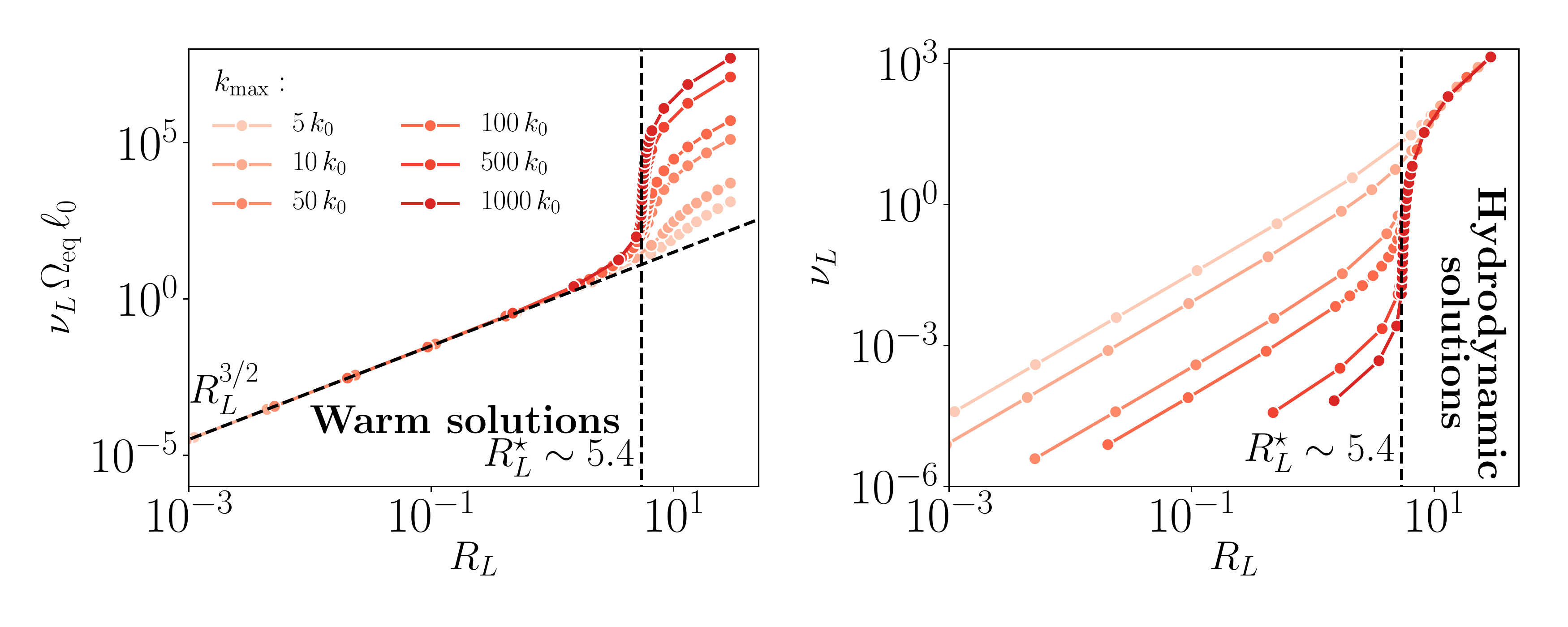}
	\put(-450,10){\large{\bf(c)}}
	\put(-190,10){\large{\bf(d)}}
	
	\caption{{\bf  Transition between warm and hydrodynamic states in the Leith model}.  
		Top panel shows the evolution of the 
		steady profiles of (a) energy spectra and (b) energy fluxes, for fixed values of
		$k_0=1/\ell_0=10^{-2}$ and $  \kmax=1$. 
		Above $\RL^\star \sim 5.4$, the energy spectra have compact support and exhibit  
		inertial range Kolmogorov scaling.  Below $\RL^\star$, both Kolmogorov and equipartition 
		scalings coexist: This is the warm regime.  The corresponding energy fluxes are then constant throughout the scales and their normalized profile become independent of $\RL$.
		Bottom panel shows the behavior of the RL viscosity as a function of $\RL$ for different values of $\kmax$: (c)  with a suitable normalization  illustrating the warm asymptotics  $\nu_L \sim R_L^{3/2}/\Omega_{\eq} \ell_0$; (d) 
  		 without normalization, illustrating universality of the hydrodynamic regime with respect to $\kmax$.
	}
	\label{fig:RLMexample}
\end{figure*}

Figure~\ref{fig:RLMexample} provides a qualitative overview of the various statistical regimes 
of the RL system that are generated by using our algorithm, which depend on the values of $\RL$.
To provide an illustrative example, we take $k_0=10^{-2}$ and $k_{max}=1$.
In \figref{fig:RLMexample} (a) and (b) we show the energy spectra and the corresponding 
energy fluxes, respectively, for various values of $\RL$.
It is clear that the RL dynamics exhibits a transition similar to the one observed for the
RNS system. Moreover, RL system refines the overall picture, as it allows us to have a 
broader range of scales and also because of its intrinsic simplicity compared to the RNS system.

At large values of $\RL$, the energy spectrum has compact support in $k$-space and is therefore of hydrodynamic type. For $\RL \simeq 29.3$, the highest values considered, the energy spectrum $E(k)$
lacks any scaling region and is concentrated around the smallest wave number $k_0$.
This is also true for the energy flux. This corresponds to an over-damped regime, wherein
the non-linear terms are unable to transfer the injected energy to smaller length-scales.

When $\RL$ is decreased ($\RL \gtrsim 5.3$), the solutions develop an inertial range with 
Kolmogorov energy spectra $E(k) \sim k^{-5/3}$, followed by a fast exponential decay to 
zero at high wave numbers;  for $\RL$ Kolmogorov scaling is present almost over the entire range
of wave numbers. This is accompanied by the associated energy flux becoming essentially constant
over an increasing range of wave numbers.
Let us remark that the development of a pure Kolmogorov scaling range in this  hydrodynamic regime  is less obvious than it would seem at first thought:
In particular, this is in contrast with the anomalous  scaling solutions that typically 
appear in freely decaying infinite-range Leith models, whose infra-red scaling exponents 
are known to be systematically larger than the constant-flux exponents \cite{connaughton2004warm,thalabard2015anomalous}.

At $\RL\sim 5.3$ the statistical character of the RL dynamics sharply changes to that of a warm 
regime. For $0 < \RL \lesssim 1.5$, both the Kolmogorov scaling  $E(k)\sim k^{-5/3}$ and 
the thermalization scaling  $E(k) \sim k^{2}$ coexist on the energy spectrum. Note that
as $\RL$ decreases down to $0$, the thermalization gradually invades the entire available 
wave number range. The associated warm energy fluxes are non-vanishing and mostly constant,
irrespective of the size of the thermalized wave number region on the energy spectrum. 
This exactly mimics the counter-intuitive behavior observed in the RNS system 
(see~\figref{fig:avgfluxdifff0})
\footnote{Similar  feature is also observed in simulations of the truncated NS equations in the vanishing viscosity limit \cite{alexakis2019energy}. This result  came to our knowledge during the revision of the present  work.}. \\

In the Leith case, this puzzling but robust signature of the warm regime can however be accounted for.  
Let  us first note that at fixed $k_0$ and $E_0$, the definition of $\RL$ implies that
$\RL \to 0$ as $\epsilon_0 \to 0$. 
We caution that the apparent constant nature of the \emph{normalized} energy fluxes
in \figref{fig:RLMexample} (b) in the $\RL \to 0$ limit should not misguide the
reader into thinking that the flux is non-vanishing in the limit $\RL \to 0$.
We emphasize that this is not the case here; this non-anomalous feature is 
in due agreement with the statement that $\RL \to 0$ corresponds to  fully 
thermalized statistics.
We also remark that in the limit $\RL \to 0$, the reversible viscosity $\nu_L \to 0$, 
as is clear from Figs.~\ref{fig:RLMexample} (c) and (d).

The fact that the flux is constant across the scales is then a direct consequence of the 
stationarity condition Eq.~(\ref{eq:LMconstEsteady}), which reduces to  $\partial_k\pi(k)=0$ 
in the limit of vanishing reversible viscosity $\nu_L$.
Solving for the corresponding energy profiles up to an overall normalization constant, we obtain
 \begin{equation}
	E(k) \propto \left(5 %
	\left(\dfrac{k}{k_{\th}}\right)^{3} %
		+  6 \left(\dfrac{k}{k_{\th}}\right)^{-5/2}%
			\right)^{2/3},
\label{eq:warmprofile}
\end{equation}
where $k_{\th}$ is the  wavenumber at which $E(k)$ reaches its minimum; 
note that $k_{\th}$ depends on the initial conditions $(\epsilon_0,k_0)$.
It is readily checked that $ E(k) \sim k^{-5/3}$ for $k\ll  k_{\th}$ and $E(k) \sim k^2$ for $k\gg  k_{\th}$:  Eq.~\eqref{eq:warmprofile} provides an explicit example of a RL warm solution!
There is therefore no contradiction in the simultaneous occurrence of a constant flux and
a warm spectrum.
Although the diffusion approximation is not valid in the RNS system, we believe that 
the property will carry through the RNS equations.  
Even though counter-intuitive, the constant fluxes observed at low $\Rr$  in \figref{fig:avgfluxdifff0} are fully compatible with the statement that the RNS statistics are warm in that regime.

\subsubsection{Reversible viscosity and the continuous phase transition}

The above description of the statistical regimes of the RL dynamics, at fixed values of $(k_0,\kmax)$,  
holds in general and is mostly insensitive to the specific choice of $\kmax$, 
provided that $\kmax/k_0$ is sufficiently large.
This can be directly inferred by monitoring  the behavior of the reversible viscosity as a function of $\RL$,  which we show in \figref{fig:RLMexample} (c) and (d) for different values of $\kmax$.
The  partitioning between the warm and the hydrodynamic regimes is clear and the transition value to
a first order approximation is independent of $\kmax$.
The sharpening of the reversible viscosity profile 
$\left. \partial_{\RL} \nu_L\right|_{\RL=\RL^\star} \to \infty$
in the thermodynamic limit $\kmax \to \infty$ suggests that the system undergoes a genuine 
phase transition at $\RL=\RL^\star \simeq 5.4$. 
The observed continuity of the reversible viscosity at $\RL \approx \RL^\star$  indicates the phase transition is  continuous.
The warm and hydrodynamic regimes can therefore be identified as genuine \emph{thermodynamic phases}.\\

In the warm phase $\RL < \RL^\star$,  the scaling behavior of $\nu_L$ for finite $\kmax$, 
is easily deduced from the definitions Eq.~\eqref{eq:LeithViscosity} and Eq.~\eqref{eq:RLparam}. 
As $\RL \to 0$, we indeed  obtain $\nu_L \sim \epsilon_0/\Omega_{\eq} \sim (E_0\RL)^{3/2} /(\ell_0 \Omega_{\eq})$, with $\Omega_{\eq}$  denoting the value of the enstrophy when the energy spectrum is fully thermalized and $\propto k^2$. \figref{fig:RLMexample} (c) shows that the scaling in fact extends up until $\RL \overset{<}{\to} \RL^\star$.
Please observe that the  dependence of $\nu_L$ on the cutoff parameter through $\Omega_{\eq} \propto \kmax^2$ signals partial thermalization of small length-scales, and implies that $\nu_L \to 0$ as $\kmax \to \infty $.

In the hydrodynamic phase, the reversible viscosity is independent of $\kmax$ (see \figref{fig:RLMexample} (d)). 
This reflects the fact that the statistics observed at finite $\nu_L>0$ are mostly 
independent of $ \kmax$ as $\kmax \to \infty$.
The continuity of the reversible viscosity implies $ \nu_L \to 0$ as $\RL \overset{>}{\to} \RL^\star$.
\subsubsection{Mean-field Landau free energy}
\label{sec:landau}
\begin{figure*}[htb]
	\includegraphics[width=1\textwidth,trim=0.5cm 1cm 1cm 0cm,clip]{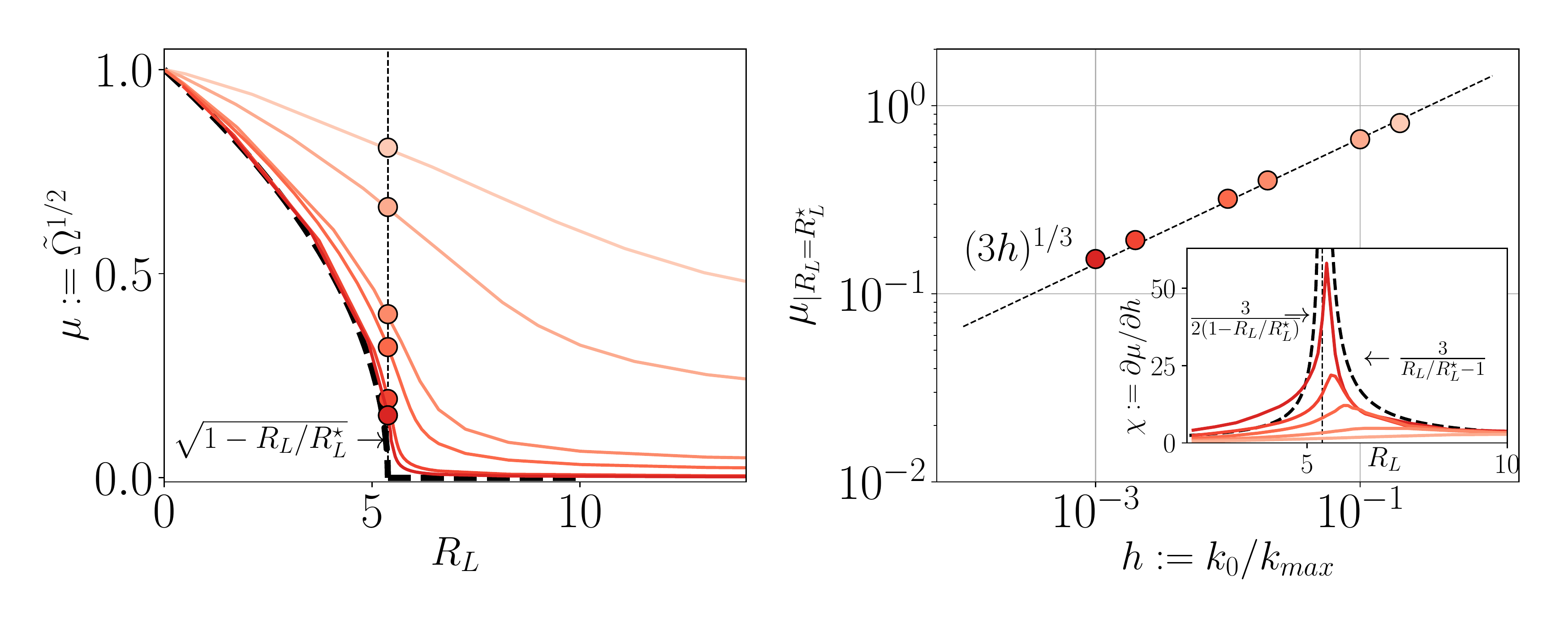}
 	\put(-450,10){\large{\bf(a)}}
	\put(-190,10){\large{\bf(b)}}
	\caption{{\bf  Mean-field behavior of  the RL  transition}.
	Panel (a) shows the spontaneous RL magnetization $\mu$ as a function of $\Rr$ for $h=k_0/\kmax$ ranging from $0.2$ to $0.001$.
	The color scale is indicated by the color of the dots in Panel (b).
	In Panel (b), the main figure shows the  RL magnetization at critical point $\mu(r=0,h)$ as a function of $h$. The inset shows the corresponding susceptibility $\chi(r,h)$ near the critical point estimated through finite-difference scheme. No estimates are provided for	$h=0.2$.
	The dashed lines indicate the  mean-field predictions Eq.~\eqref{eq:LandauPred}.
 	}
	\label{fig:RLMeanField}
\end{figure*}
The interpretation of the RL ``Warm/Hydrodynamic'' transition in terms of a continuous phase transition can in fact be further substantiated and  described in terms of a heuristic Landau theory (see, \emph{e.g.}, \cite{goldenfeld2018lectures}) involving the parameters 
\begin{equation}
	\begin{split}
		&r:= \RL/\RL^\star-1 \hspace{0.3cm} \text{(``reduced temperature'')}, \\
		&\mu:=  \tilde \Omega^{1/2}=\left( \Omega/\Omega_{\eq}\right)^{1/2} \hspace{0.3cm} \text{(``magnetization'')},\\
		&h := k_0/\kmax \hspace{0.3cm} \text{(``magnetic field'')}.
	\end{split}
	\label{eq:LandauParam}
\end{equation}
Our choice of parameters is essentially data-driven and motivated by a succession of empirical trials.
Note that the order parameter $\mu$ can as well be defined using the reversible viscosity as $\mu \propto  \nu_L^{-1/2}$.  This hydrodynamic analogue of magnetization is sensitive to ultra-violet thermalization and takes value $1$ at $\RL =0$. In the hydrodynamic phase, it  is non vanishing  for finite $\kmax$, but  converges to $0$ at $\RL^\star$ as $\kmax \to \infty$.
This feature together with the fact that the transition is smooth (absence of singular behavior) for finite values of $\kmax$ is the main indication as to why (one over) $\kmax$ may be suitable as a smoothing symmetry breaking parameter.
Let us now consider the mean-field Landau free energy
\begin{equation}
	\begin{split}
	&\phi_L(\mu,h, r):= -3 h \mu + \dfrac{1}{2}r^2 \mu^2 + \dfrac{1}{4} \mu^4, \\
	&\text{defined for } \mu \ge 0, h \ge 0, r \ge -1.
	\end{split}
	\label{eq:Landau}
\end{equation}
In spite of its simplicity, \figref{fig:RLMeanField} reveals that  the mean-field free energy $\phi_L$  captures the essential signatures of  the RL transition between the warm and hydrodynamics phases.
In particular, it predicts the following magnetization profiles
\begin{equation}
	\begin{split}
	&\mu(r,h=0^+) = \sqrt{1 -r} \hspace{0.3cm} \text{if $ r>0$}  \hspace{0.1cm}\text{ and $0$ otherwise},\\
	&\mu(r=0,h) = (3h)^{1/3},\\
	&\chi(r,h) := \dfrac{\partial \mu}{\partial h } = \dfrac{3}{r} \hspace{0.2cm} \text{if $r>0$}  \hspace{0.1cm}\text{and $\dfrac{3}{2 r} $ otherwise}.
	\end{split}
	\label{eq:LandauPred}
\end{equation}
\figref{fig:RLMeanField} shows that the lowest order predictions for the magnetization are in excellent agreement with the RL data.
The spontaneous magnetization $\mu(r,h)$ indeed seems to converge towards the mean-field prediction as $h$ is decreased towards $0$.
At the critical point, the scaling with $h$ is close to perfect over two decades. This suggests that as $h \to 0$, the susceptibility $\chi(r=0,h)$ indeed genuinely diverges, and that the RL transition is
a continuous phase transition.  However, we notice
deviations from our  mean-field prediction for the susceptibility. 
The mean-field exponent is compatible with the data, but a finer assessment would require reaching higher resolutions. 
We remark that the behavior of $\chi(r,h)$ in the hydrodynamic phase is seemingly 
independent of $h$, as predicted by Eq.~\eqref{eq:LandauPred}.
This is not the case in the warm phase and results in deviations from our heuristic mean-field predictions.\\

Our  partial conclusion  at this point  is that in spite of the warm second-order deviations from the mean-field, \figref{fig:RLMeanField}  suggests that  the RL transition between warm and hydrodynamic states indeed fits into a general thermodynamic framework  and corresponds to a continuous phase transition.  It is unclear whether specific properties of the transition could be deduced from first principles, but these considerations are beyond the scope of the present work. 
\subsection{From  Reversible-Leith to  Reversible Navier-Sokes dynamics}

Even though the RL  model has a very simple dynamics, it is naturally tempting 
to infer that the RNS statistics  fits into a similar general thermodynamic framework as the one identified in the Leith case. We shall not refrain from doing so in \S~\ref{sec:RLtoRNS}, in order to explore a thermodynamic formulation of  Gallavotti's conjecture that could have practical implications for its  rigorous numerical assessment. Prior to this, let us however briefly point out some important differences between the RL and the RNS formulations.\\

\paragraph{$\RL$ \emph{vs.} $\Rr$.}\hspace{0.1cm}\\
The specific definition of the reversible control parameter is different in the RNS and RL 
formulations, being dependent on the reversible dynamics in the respective systems.
Therefore,  we do not  expect the RL free energy Eq.~\eqref{eq:Landau} to account for the finite-size effects observed in our RNS numerics.
We can investigate the RNS system in terms of the following redefined control parameter
\begin{equation}
\label{eq:RNSRtilde}
\tilde \Rr:= \av{\epsilon_\inj}^{2/3} \ell_f^{2/3}E_0^{-1}; 
\end{equation}
this  would provide an exact analogue of the definition Eq.~\eqref{eq:RLparam}.  
This definition is more appropriate for a comparative study of the RNS system under different
forcing schemes. 
The only drawback of such a definition is that it relies on a data-driven measurement, namely that of $\av{\epsilon_\inj}$. 
In the present work, we see this only as a minor issue, with no relevance for the forthcoming discussion. 
In our view, this is one of the reason why the RNS mean-field description fails to account for the
RNS statistics in a strict sense. However, upon defining the RNS hydrodynamic magnetization as $\mu_{\rm RNS} := \tOm$ rather than the square-root  RL definition Eq.~\eqref{eq:LandauParam}, and using  $r_{\rm RNS} := \Rr/\Rr^\star -1$ as the RNS reduced temperature, it is apparent that the mean-field free-energy prescribed by Eq.~\eqref{eq:LandauParam} correctly accounts for the phenomenological square-root fit of \figref{fig:globquant}. Note that higher-resolution runs are needed to examine the role of $\kmax$ on
the RNS transition.

\paragraph{Dynamical behavior within the transition region.}\hspace{0.1cm}\\
Another difference between the present RNS and  RL formulations relates to the fact that 
the injected power  fluctuates in time in the former.  
We think that this  feature prevents the RL dynamics from having a noticeable transition region 
at finite $\kmax$. Specifically in the Leith model, for a given input flux $\epsilon_0$ and prescribed finite $\kmax$  only one 
steady solution exists. It is either a warm solution or a hydrodynamic solution:  In other words there is no multi-stability of solutions.
This could well be the case in the RNS system, if the injected power was constant in time.
However, in our RNS formulation the injected power $\epsilon_\inj$ is 
constant only on average.  For the fixed total energy, the input parameter $f_0$ only 
imposes the upper bound:    
$|\epsilon_{\inj}|\le 2 E_0 |f_0|$. The  enstrophy fluctuations observed in 
the transition region for the  RNS simulations  could therefore be simply due to power fluctuations,  selecting either a warm solution or a hydrodynamic solution as a function of the instantaneous value of $\epsilon_\inj$.
These fluctuations  therefore are not captured by  the  mean-field Landau description Eq.~\eqref{eq:LandauParam}. Further intuitions on this matter could  perhaps be obtained by   generalizing  the deterministic RL framework to a stochastic one, 
but this is beyond the scope of the present work.\\

It is insightful to use the RL analysis as a guiding framework, irrespective of the 
intrinsic differences between the RNS and RL systems, to interpret the RNS transition as
a genuine continuous phase transition in the limit $\kmax \to \infty$. 
This  has  practical implications for the Gallavotti's \emph{equivalence conjecture}, 
which we discuss in the next section.

\section{Turbulent limit, critical point and Gallavotti's conjecture}
\label{sec:RLtoRNS}
So far, we have focused on the RNS statistics, without any reference to the standard NS statistics, 
except for some very qualitative comments.
However, as explained in the introduction, the motivation in studying reversible dynamics
is to assess whether  Gallavotti's \emph{equivalence conjecture} holds true.
In essence, the conjecture states an identity between the RNS and NS invariant measures,  
hereafter denoted as $\av{\cdot }_{E_0}$ and $\av{\cdot}_\nu$, respectively.
We quote directly from Ref.~\cite{biferale2018equivconj} and state the equivalence
as an (asymptotic) statistical identity valid for a suitable class of observable $\mO$, \emph{e.g.}
\begin{equation}
	\label{eq:equivalence}
	 \av{\mO}_\nu  = \av \mO_{E_0}\,(1+\circ(1)),
\end{equation}
where 	$\circ(1)$ denotes a vanishing quantity in a suitable joint limit $\nu\to0, \kmax \to \infty$ 
\footnote{Asymptotic statistical identities such as \eqref{eq:equivalence} are exactly the kind involved in  modern treatments of the   equivalence between the canonical and the micro-canonical statistical ensembles, known to be  valid for a wide class of systems in equilibrium statistical mechanics (see, \emph{e.g.}, \cite{touchette2015equivalence} and references therein).}.
We will no further comment  on the notion of a ``suitable class of observables", except that the latter must contain the energy $E$, so that the following \emph{reflexivity property} holds:
\begin{equation}
	\label{eq:reflexivity}
	 \av{E}_\nu  = E_0\,(1+\circ(1)).
\end{equation}
The notion of a ``suitable joint limit'' $\nu\to0,~\kmax \to \infty$, 
however needs further substantiation.
In Ref.~\cite{biferale2018equivconj}, the authors  consider the limit $\nu \to 0$ at fixed value of $\kmax$. This limit is particularly relevant in the perspective of the many interesting recent developments related to Galerkin-truncated dynamics \cite{kraichnan1975remarks,cichowlas2005effective,frisch2008hyper,krstulovic2008two,krstulovic2009cascades,ray2015thermalized}.
In this limit, it is now known that the NS equations generate quasi-equilibrium flows \cite{alexakis2019energy}. It is therefore  not the relevant asymptotics  in the context of describing fully developed turbulence, which  the (truncated) NS equations in principle generate in the joint ordered limit 
\begin{equation}
\label{eq:turblim}
 \kmax \to \infty \,\,\text{ first},   \,\,  \nu \to 0 \,\,\text{ then},
\end{equation}
 at fixed forcing statistics. Within the turbulent limit \eqref{eq:turblim}, the statistical identity  \eqref{eq:equivalence} describes a dynamical equivalence between the \emph{full} RNS and the \emph{full} NS statistics. To the best our understanding,  this also corresponds to the original formulation of the equivalence conjecture \cite{gallv96PLAequivconj,gallaPhysD97}. \\

Let us now explicitly assume that the RNS steady statistics are  described by 
a continuous phase transition at $\Rr^\star$ as $h:=k_0/\kmax \to 0$.  
One can then precisely identify the turbulent limit as the critical point asymptotics, 
approached from the hydrodynamic phase.

\begin{figure*}[ht]
\includegraphics[width=\textwidth,trim=0.5cm 1cm 1cm 0cm,clip]{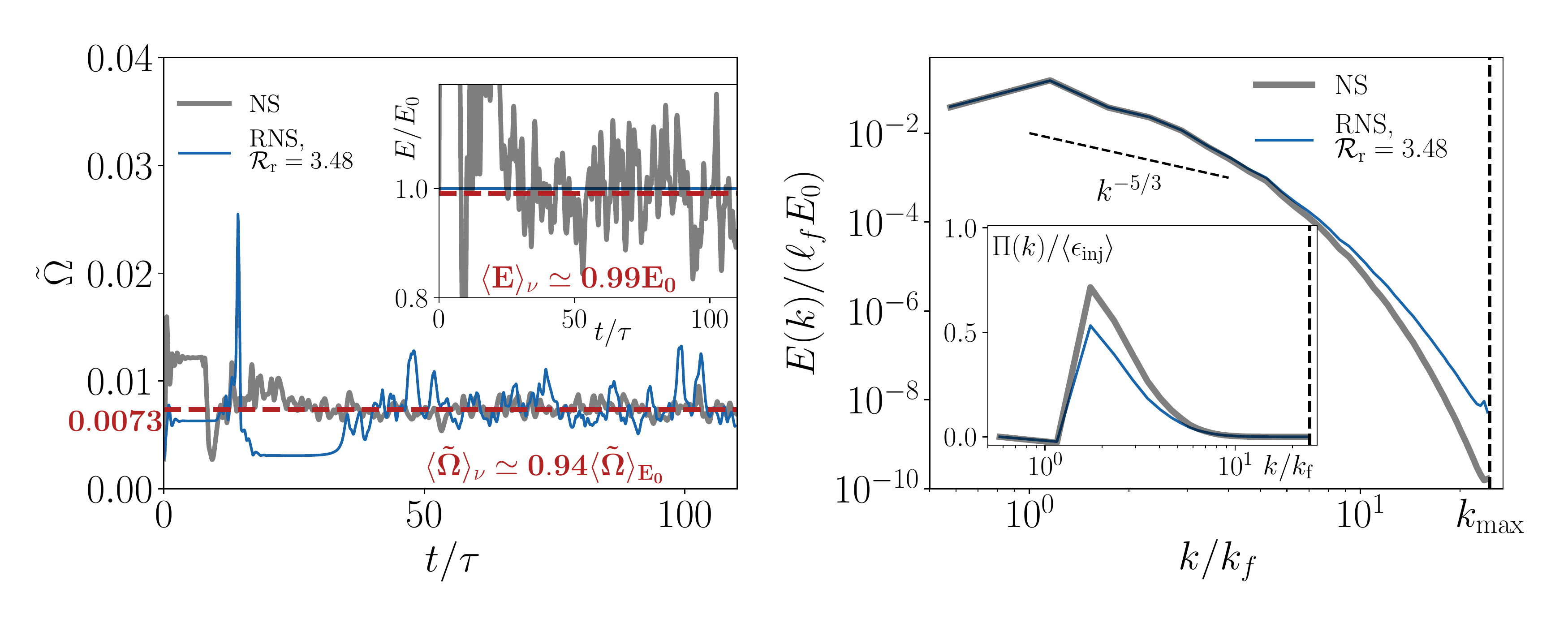}
\put(-460,10){\large{\bf(a)}}
\put(-200,10){\large{\bf(b)}}
\caption{
{\bf  RNS \emph{vs.} NS  at $\Rr  \gtrsim \Rr^\star$. }
	Panel (a) superimposes the times series of the normalized enstrophy (main panel)  for  the \Aone\, RNS run  at $\Rr =3.48$ and a corresponding NS run, with same Taylor-Green forcing   and standard viscosity set to $\nu=\av \nur$.  Inset shows the corresponding data for the energy time series, showing the validity of the reflexivity property (see text). The dashed-red lines indicate  NS time-averages. Panel (b) compares the  time-averaged energy spectra (main panel) and fluxes (inset)
for both the RNS and NS runs.  Kolmogorov scaling is indicated by a dashed line with label $k^{-5/3}$. $E_{0}$ is the conserved total energy of the RNS run.
}
\label{fig:globRNSNS}
\end{figure*}
\subsection{Turbulent limit and critical point asymptotics: $\Rr \to \Rr^{\star} \,\,\& \,\,h \to 0$.}
\label{sec:TurbulentLimit} 
The identification of the turbulent limit as the critical point asymptotics is a consequence of the  transition being continuous. 
Indeed, recalling the  warm  behavior $ \av{\nur} \propto h^{2} \to 0$ illustrated in \figref{fig:globquant}~(a) and \figref{fig:RLMexample}~(c), we infer that the reversible viscosity is uniformly vanishing for $\Rr<\Rr^\star$ in the  limit $h\to0$.
Assumed continuity of the transition then yields $\av{\nur} =0$ at $\Rr=\Rr^\star$.
Besides, for example, assuming a constant energy $E_0$  as $\Rr \to \Rr^\star$, the forcing amplitude converges towards a finite limit, so that  
$f_0 \to f_0^\star := \Rr^\star E_0/\ell_f$: We have recovered the  turbulent limit~\eqref{eq:turblim}.\\

We comment on two salient features of this thermodynamic reformulation of the RNS turbulent limit.
\paragraph{Order of the limits: }\hspace{0.1cm\\}
In the limit $h\to 0$, the warm states are strictly speaking ill-defined as a consequence of the ultra-violet catastrophe, as partially explained in  Appendix~\ref{sec:abseq}.
Hence,  approaching the critical point from below necessarily requires taking the limit
 $\Rr \overset<{\to} \Rr^\star$  \emph{before} $h\to0$.
In contrast, the hydrodynamic states are well defined even as $h\to0$: We have extensively argued throughout our exposition that the statistics are independent of the cutoff in this phase. We therefore conjecture that approaching the  critical point from above can therefore also be done by taking the thermodynamic limit  \emph{before} the critical limit; in other words, the limits $h\overset >\to0$, and $\Rr\overset>\to \Rr^*$ should in principle commute. We can therefore unambiguously refer to the (unordered) joint limit $h \overset>\to0,\, \Rr\overset >\to \Rr^*$ as the ``turbulent limit''.\\

\paragraph{Anomalous dissipation:}\hspace{0.1cm\\}
In the  limit $h~\overset{>}{\to}~0,\, \Rr\overset{>}{\to}~\Rr^*$, we could in principle expect  anomalous dissipation from the RNS statistics. This is yet better seen if the  alternative definition $\tilde \Rr$ of Eq.~\eqref{eq:RNSRtilde} indeed could be used as a valid  reversible control parameter. We will then  obtain $ \av{ \epsilon_\inj} \to \epsilon^\star := \ell_fE_0^{3/2} {\tilde{\Rr^\star}}{} ^{3/2} <\infty$. This argument suggests that the scale-by-scale energy budget and the associated 4/5 laws  could be deduced following the exact same steps as for   the standard NS equations \cite{frisch1995turbulence}, hinting at the equivalence between those two dynamics at the critical point. 

\subsection{RNS \emph{vs.} NS near criticality: Illustrative numerics }
\label{sec:TurbulentLimitIllus} 
As a final illustration of the relevance of the   joint limit $\kmax \to \infty,\, \Rr\overset> \to \Rr^\star,$  let us here explicitly  compare  the RNS statistics from Set \Aone\, to their NS counterpart at $\Rr \sim 3.48 >\Rr^+,$ a value which corresponds to the lower end of the hydrodynamic regime for our resolution. As stated before,  the RNS system  then produces a non-trivial statistical state, which involves multitude of length and time scales.  To generate corresponding NS steady states, we integrate the NS equations  with same Taylor-Green forcing amplitude $f_0$  and set the standard viscosity to $\nu=\av \nur$.  The main results are summarized in \figref{fig:globRNSNS}. The inset of Panel (a)  shows that  the  NS energy  fluctuates around the imposed RNS value, \emph{e.g.} $\av{E}_\nu \simeq 0.99 \,E_0$,  and this reflects the approximate validity of the reflexivity condition prescribed by Eq. \eqref{eq:reflexivity}. The main panel shows that the NS and the  RNS  enstrophy time-series fluctuate around a similar mean value. The fluctuations are more or less commensurate with each other (slightly larger for the RNS run).

\old
\figref{fig:globRNSNS}~(b) shows that the  RNS and the NS dynamics in this regime have in fact  similar large-scale features. In particular, both the spectra and the fluxes  show excellent agreement on a decade of wave numbers ($k < 10 \kf$), before starting to deviate in the  ultra-violet range $k > 10 \kf$. This is a consequence of our simulations having finite resolution, and these differences would probably disappear upon taking larger $\kmax$ for same $f_0$.

\old
\section{Concluding Remarks}
\label{sec:conclusions}
Time-reversible formulations of  forced-dissipative hydrodynamical equations, 
addressing Gallavotti's  equivalence conjecture of (hydro)dynamical ensembles  
have emerged in recent years as an important framework to provide an out-of-equilibrium 
thermodynamic perspective on the issue of turbulent irreversibility.
Yet, in spite of many promising recent  numerical results using reduced models, 
circumstances under which the equivalence conjecture might hold true remains unclear. 
Also, the attention has recently shifted to analyze (reversible) models wherein thermostats
preserve various other quadratic quantities, and not necessarily the total energy.
Within these frameworks, the equivalence conjecture has been assessed in the near equilibrium
regime, which corresponds to the vanishing viscosity at finite resolution~\cite{biferale2018equivconj,depietro2018shell,gallv2019private}.
In the present work, we have followed a completely different route, in order to provide 
intuition on the potential validity of the equivalence conjecture for the full 3D dynamics, 
in the limit $\kmax \to \infty,\, \nu \to 0$. To this end, our analysis has focused on
studying the dynamics of the RNS system that preserves the total kinetic energy.  
We carried out an extensive numerical study of the RNS system to fully explore its 
statistical regimes and also provide an illustrative comparison of the RNS and NS statistics.
We find this approach particularly insightful.

Our numerics show that the RNS system undergoes a continuous phase transition controlled by
a non-negative dimensionless parameter $\Rr=f_0\ell_{\rm f}/E_0$, which quantifies the 
balance between the fluctuations of kinetic energy at the forcing length-scale $\ell_{\rm f}$ 
and the total energy $E_0$; $f_0$ is the forcing amplitude.
In our opinion, it is possible to use an alternative definition (data driven) of the control 
parameter, without modifying the overall picture: 
$\tilde \Rr = \av {\epsilon_\inj}^{2/3}\lf^{2/3}/E_0$,
where $\av {\epsilon_\inj}$ is the stationary value of the injected power.
For small $ \Rr$, the RNS dynamics produces a ``warm'' stationary statistics, 
\emph{e.g.} characterized by the partial thermalization of the small length-scales and an 
intrinsic dependence on the cutoff  $\kmax$.
At large $\Rr$, the stationary solutions have  a hydrodynamic behaviour, characterized 
by compact energy support in $k$-space and are essentially insensitive to  the 
truncation scale $\kmax$.   \\

The transition between the two statistical regimes is observed to be smooth, with a
narrow crossover range in the vicinity of $\Rr^\star \simeq 2.75$. It is characterized by a
highly bursty dynamical behavior of the enstrophy, whose fluctuations are commensurate with 
its mean. In this regime, the system exhibits multi-stability: It oscillates between a 
hydrodynamic-type low-enstrophy regime and a high-enstrophy regime, whose small length-scale 
statistics are yet far from being thermalized and exhibit  non-trivial power law scalings. 
The enhancement of the enstrophy fluctuations in this transition region hints at a continuous 
phase transition between the warm regime and the hydrodynamic regime, with time-averaged
reversible viscosity (or equivalently normalized enstrophy) emerging as an order parameter.
The transition in principle occurs only in the thermodynamic limit $\kmax \to \infty$. 
Therefore, a strict characterization of the transition must necessarily involve 
finite-size scaling analysis.

\begin{figure}[bt]
\includegraphics[width=\columnwidth,trim=0.5cm 1cm 0.5cm 0cm, clip]{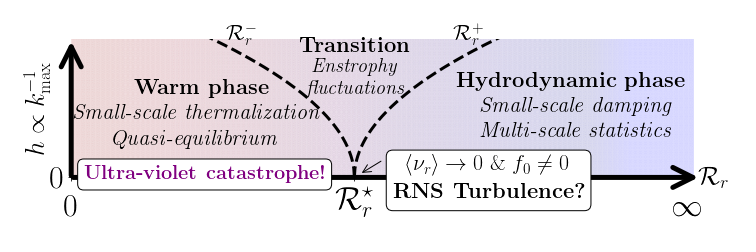}
\caption{A refined phase diagram for the RNS steady-state. 
Our numerical simulations suggest $\Rr^\star \approx 2.75$.   }
\label{fig:Sketch2}
\end{figure}

To further substantiate this idea, we used a  simple one-dimensional non-linear 
``Leith-type'' diffusion model, modified to preserve energy, so that it mimics the RNS system. 
The main difference between the RL and the RNS system is the forcing scheme, which in the former is imposed by requiring constant energy fluxes at the boundaries.
RL formulation allows the computation of steady states without relying on the direct 
numerical integrations of the RL equations, but rather using a non-trivial parametrization 
and ideas from the theory of dynamical systems.
Similar to the RNS analysis, the RL steady regimes were classified depending on  
a dimensionless control parameter  $\RL$ and these mimic the smooth transition between the 
warm and hydrodynamics states. \\

The simplicity of the RL formulation allowed us to investigate in detail the finite-size effects 
and the related influence of the cutoff $\kmax$. 
This asymptotic analysis substantiated the idea of a continuous phase transition: 
In fact, we find that the signatures of the phase transition close to the critical 
point $\Rr^\star$ can be obtained by constructing a heuristic mean-field Landau free energy.
In this picture,  $\Rr$ indeed behaves as a thermodynamic control parameter, \emph{e.g.} a temperature;
the relevant order-parameter is defined in terms of a suitably normalized enstrophy, 
while the symmetry breaking parameter $h$ is identified as the cutoff length-scale $1/\kmax$.\\

\old
Naturally, the RL dynamics only reproduces the idealized features of the RNS transition 
and differs from the RNS system in important ways.
In the Leith model, the critical control parameter is exclusively identified from 
the properties of the average steady state, and therefore it does not account for the 
dynamical signatures  of the transition that are found in the RNS system,  
namely  the  enhancement of enstrophy variance in the vicinity of the critical point. 
Also, the high-enstrophy region ($\RL<\RL^\star$) exhibits close to thermalization spectra
$E(k)\sim k^2$ up until $\RL^\star$; this is in contrast with the RNS observations, wherein
close to the critical value $\Rr^\star$, the power-law exponents of the energy spectra
at small length-scales are observed to fluctuate, but are bounded by $2$.
This signals a clear departure from the Gibbsian equipartition in the RNS system, 
and this is not captured by the simplified model. 
In our view, these differences can for the most part be traced back to the fact that 
the injected energy fluctuates in the RNS system, but is held constant in the RL model. \\

In spite of the above differences, it is quite evident that the RNS systems exhibits a
continuous phase transition; therefore, the phase diagram is qualitatively 
similar to the one obtained for the RL model (\figref{fig:Sketch2}).
This makes it possible to examine the  Gallavotti's equivalence conjecture by formulating
the turbulent limit in terms of the critical point
asymptotics $\Rr \overset>\to \Rr^\star , h\overset>\to 0$, with the overset symbol ``$>$'',
where limit is approached from the hydrodynamic regime.
We have argued that in this limit the RNS states should have anomalous energy dissipation
and formally vanishing thermostat effects, thereby hinting at the validity of the conjecture;
a comparison of the RNS and NS numerics indeed suggests that this is true.
Moreover, we strongly believe that the limits  $\Rr \overset>\to \Rr^\star$ and
$ h\overset>\to 0$ commute, compared to the standard formulation of the turbulent 
limit as $\kmax\to \infty, \nu \to 0$ in the NS equations, this will then constitute a
major simplification, and hopefully will pave the way for systematic assessment of the
equivalence conjecture in future studies.

\old
In the present work, we have restricted ourselves to DNSs involving 
rather modest numbers of grid points (up to $N^3_c=128^3$) and our description is
based on one point statistical quantities. 
The complete characterization of the statistical regimes of the RNS system with larger
grid sizes is computationally very demanding, given that many of the runs require long
temporal evolution. Yet, our results suggest to study the asymptotic behavior of RNS 
only at the transition, namely by letting $N_c \to \infty$ at fixed $\Rr \gtrsim\Rr^{\star}$. 
While we have provided the evidence that in this regime the RNS system correctly reproduces the 
macroscopic properties of the NS equations,  a systematic asymptotic analysis 
is still desirable, to investigate the nature of the  agreement at higher Reynolds number. 
Moreover, a careful investigation of more refined statistical properties, beyond the 
relatively low-order statistics considered here, is needed to complete the picture, 
but we leave it for future investigations.\\

\paragraph*{Acknowledgments.}
Part of this work was granted access to the following computing grants: 
GENCI (Grand Equipement National de Calcul Intensif)  grant numbers: 
A0042A10441 (IDRIS and CINES), A0062A10441 (IDRIS, CINES and TGCC) and 2A310096 (IDRIS). 
B. Dubrulle acknowledges funding from ANR EXPLOIT, grant agreement no. ANR-16-CE06-0006-01. 
S. Nazarenko is supported by Chaire D'Excellence IDEX, Universit\'e de la C\^ote d'Azur, France.
%

%

\appendix
\section{Absolute equilibria of the truncated Euler equations}
\label{sec:abseq}
For $\Rr = 0$, the numerical integration of the RNS equations exactly reduces to integrating the so-called ``truncated Euler'' equations. These are in fact obtained by performing a Galerkin truncation 
of the Euler equation  at a cutoff wave number $\kmax$. In practice, Galerkin  truncations consist in suppressing all  the triadic  interactions involving wave numbers larger than $\kmax$, whereby yielding 
a high-dimensional conservative set of non-linear ordinary differential equations.
Truncated Euler flows exactly preserve the quadratic invariants of the original equations and satisfy a Liouville theorem.
Hence, they typically  converge towards  thermal statistical states with Gibbsian statistics  known as ``absolute equilibria''~\cite{lee1952some,kraichnan1967inertial,kraichnan1973helical}, and the  thermalization process usually exhibits  interesting transients~\cite{cichowlas2005effective,krstulovic2008two,krstulovic2009cascades}. 
For the non-helical 3D truncated Euler flows that we consider in the present paper, the relevant absolute equilibrium state is particularly simple and prescribed by each Fourier velocity mode having independent centered Gaussian statistics   with variance $\propto E_0/N_c^3$. This equilibrium state describes an equipartition of the total kinetic energy $E_0$ among the different modes.
Assuming a continuous distribution of wave numbers,  the corresponding energy spectrum  can  be estimated as
\begin{equation}
	E_{\eq}(k)=\frac{3 E_0}{\kmax^3} k^2,
\end{equation}
 and  the  absolute equilibrium enstrophy is then 
\begin{equation}
	\label{Eq:OmegaEQabs}
	\Omega_{\eq}:=\int_1^{\kmax}k ^2 E_{\rm eq}(k)\, {\rm d}k \underset{\kmax \to \infty}{\sim} 
	\frac{3}{5}E_0 k_{\rm max}^2.
\end{equation}

In the limit  $\kmax \to \infty$, the absolute energy equilibria become ill-defined, as the resulting energy spectra cannot be normalized, unless they are trivial and the total energy is vanishing. 
This phenomenon  is the so-called ``ultra-violet catastrophe'', which also prevents the warm RNS states  to be properly defined  in the limit $\kmax \to \infty$.

\end{document}